\documentclass[superscriptaddress,amsmath,amssymb,aps,prd,preprintnumbers,showpacs,twocolumn,nofootinbib]{revtex4-2}

\usepackage[T1]{fontenc}
\usepackage[utf8]{inputenc}
\usepackage{amsmath,amssymb,bm,mathtools,natbib}
\usepackage{graphicx}
\usepackage[caption=false]{subfig}
\usepackage{slashed}
\usepackage{dsfont}
\usepackage{float}
\usepackage{cancel}
\usepackage{units}
\usepackage{array}
\usepackage{upgreek}
\usepackage{booktabs}
\usepackage[dvipsnames,table,xcdraw]{xcolor}
\usepackage{enumerate}
\usepackage[normalem]{ulem}
\usepackage[colorlinks=true,pdfstartview=FitV,linkcolor=blue,citecolor=blue,urlcolor=blue,breaklinks=true]{hyperref}
\usepackage{orcidlink}

\begin{document}

\title{Thermodynamics and emergent thermomechanical response of a quantum ring \\ with nonminimal spin--orbit coupling}

\author{Jo\~{a}o A.A.S.\ Reis\orcidlink{0000-0002-2831-5317}}
\email{joao.reis@uesb.edu.br}
\affiliation{Departamento de Ci\^{e}ncias Exatas e Naturais, \\ Universidade Estadual do Sudoeste da Bahia, Itapetinga (BA), 45700-000, Brazil}

\author{L. Lisboa-Santos\orcidlink{0000-0003-4939-3856}}
\email{leticia.lisboa@discente.ufma.br}
\affiliation{Programa de P\'os-Gradua\c{c}\~{a}o em F\'{\i}sica, Universidade Federal do Maranh\~{a}o, Campus Universit\'{a}rio do Bacanga, S\~{a}o Lu\'{i}s (MA), 65085-580, Brazil}

\author{Edilberto O. Silva\orcidlink{0000-0002-0297-5747}}
\email{edilberto.silva@ufma.br}
\affiliation{Departamento de F\'{\i}sica, Universidade Federal do Maranh\~{a}o, 65085-580 S\~{a}o Lu\'{\i}s, Maranh\~{a}o, Brazil}

\date{\today}

\begin{abstract}
We investigate the thermodynamic and emergent thermomechanical properties of fermions confined to a one-dimensional quantum ring with effective spin--orbit interactions induced by nonminimal couplings to antisymmetric tensor fields. Using the exact spectrum obtained in the companion work, we develop canonical and grand-canonical descriptions and show that the coupling parameter~$\xi$ deforms the angular-momentum branches, reorganizing the low-energy spectrum and leaving clear signatures in the internal energy, entropy, heat capacity, and spin--orbit response functions. We also formulate an effective thermomechanical description by treating the ring circumference as a quasi-static thermodynamic variable. This leads to a pressure-like quantity and associated response coefficients, directly linked to the microscopic spectrum. In the grand-canonical ensemble, Fermi statistics strongly enhance the response, producing coupling-dependent instabilities and sign changes reminiscent of mesoscopic de~Haas--van Alphen oscillations. Finally, we introduce a phenomenological interacting extension based on an exponential resummation of the free energy, showing that collective effects can sharpen the thermomechanical response and induce anomalous thermal contraction. Our results connect spectral deformation, finite-size thermodynamics, and emergent mechanical behavior in spin--orbit-active quantum rings.
\end{abstract}

\maketitle

\section{Introduction}
\label{Sec:Introduction}

Low-dimensional quantum systems provide a fertile ground for the
emergence of nontrivial physical phenomena driven by the interplay
between geometry, quantum coherence, and spin dynamics~\cite{Imry2002,Viefers2004}.
In particular, fermions confined to mesoscopic structures such as quantum rings
exhibit a rich variety of effects---including geometric
phases~\cite{Berry1984,AharonovAnandan1987},
persistent currents~\cite{ButtikerImryLandauer1983,Levy1990,Chandrasekhar1991,CheungGefenRiedel1988},
and spin-dependent transport~\cite{Frustaglia2004,Splettstoesser2003}---which
have been extensively explored both theoretically and experimentally
in the context of spin--orbit interactions and
topological systems~\cite{HasanKane2010,QiZhang2011}.

In conventional condensed matter settings, spin--orbit coupling arises
from relativistic corrections associated with structural inversion
asymmetry or external fields, leading to well-known models such as the
Rashba~\cite{BychkovRashba1984} and Dresselhaus~\cite{Dresselhaus1955}
Hamiltonians. The strength of the Rashba interaction can be tuned
electrically via gate voltages~\cite{Nitta1997,Grundler2000}, a feature
that has stimulated a broad program of spintronics
research~\cite{ZuticRMP2004,ManchonRMP2015,Winkler2003,SinovaRMP2015}.
More generally, however, spin-dependent interactions can also emerge from
nonminimal couplings in relativistic field theories, where fermions interact
with external fields through derivative terms~\cite{ColladayKostelecky1998,KosteleckySamuel1989,KosteleckyMewes2002,
CarrollFieldJackiw1990,Mattingly2005}.
These couplings naturally induce effective gauge structures in the
nonrelativistic regime, providing an alternative and conceptually broader
route to spin--orbit physics that connects high-energy theory with
mesoscopic condensed matter
phenomenology~\cite{BelichSilva2011,Belich2009,BakkeSilva2012,BelichDenisov2006}.

Recently, it has been shown that such nonminimal couplings, when
restricted to a ring geometry, leads to an effective one-dimensional
model with a characteristic Rashba-like
structure~\cite{SilvaReisLisboa2025,ReisSilva2024}.
In particular, the resulting energy spectrum is not obtained through a
simple splitting of degenerate levels. Instead, the effective coupling
produces a deformation of the angular-momentum branches themselves,
shifting their centers and reorganizing the ordering of low-energy
states. This mechanism is fundamentally different from the standard
Rashba splitting observed in semiconductor quantum
rings~\cite{Meijer2002,Molnar2004,ShengChang2006} and has direct
consequences for all physical properties of the system.

From an experimental standpoint, semiconductor quantum rings have been
realized in several material platforms. Self-assembled InAs quantum rings
were characterized spectroscopically by Lorke et
al.~\cite{Lorke2000}, and their energy spectra were probed by Fuhrer et
al.~\cite{Fuhrer2001}. The interplay between spin--orbit coupling and
ring geometry has been demonstrated through Aharonov--Casher
interference experiments~\cite{Koga2002,Bergsten2006,Konig2006},
and the direct control of the spin geometric phase in rings with Rashba
coupling was achieved by Nagasawa et al.~\cite{Nagasawa2013}.
These advances establish quantum rings as a mature experimental platform
in which the theoretical predictions developed here could, in principle,
be tested.

Despite these developments, most studies of such systems have focused
on spectral and transport
properties~\cite{AharonovBohm1959,AharonovCasher1984,Stern1992,NittaKoga1999,FoersterBuettiker2008},
while their thermodynamic behavior remains largely unexplored. This is especially relevant in mesoscopic systems, where the discrete nature of the spectrum and finite-size effects can lead to nontrivial thermal signatures~\cite{TanInkson1999,TanInkson1996,BakkeFurtado2012,deLiraSilva2024}.
In particular, it is natural to ask how the deformation of the spectrum induced by nonminimal couplings manifests itself in macroscopic quantities such as the internal energy, entropy, and heat capacity, and whether it gives rise to effective mechanical response functions analogous to those found in bulk thermodynamic systems~\cite{PathriaBeal2011,LandauStat}.

In this work, we address these questions by developing a systematic
thermodynamic and thermomechanical description of a fermionic system
confined to a quantum ring and subject to effective spin--orbit
interactions induced by nonminimal couplings. Starting from the exact
energy spectrum derived in our companion work~\cite{SilvaReisLisboa2025},
we construct both canonical and grand-canonical formulations and analyze
the resulting thermodynamic behavior in detail.

A central aspect of our approach is the introduction of a consistent thermomechanical framework based on a quasi-static interpretation of
geometric variations. Although the ring radius is fixed at the level of the Hamiltonian, the explicit dependence of the spectrum on the geometric parameter allows us to consider infinitesimal, adiabatic variations of the system size, in direct analogy with the treatment of lattice constants in
solid-state physics.
Within this framework, the circumference of the ring can be promoted
to an effective thermodynamic variable, making it possible to define
a pressure-like quantity and associated response coefficients.

This construction reveals that the system exhibits a well-defined
thermomechanical response, in which quantities such as the effective
pressure, isothermal compressibility, and thermal expansion coefficient
emerge directly from the quantum spectrum.
In particular, we show that the effective pressure is proportional
to the internal energy density, while the thermal pressure coefficient
is directly related to the heat capacity.
In the grand-canonical ensemble, Fermi-surface effects amplify these
signatures dramatically, producing coupling-dependent mechanical
instabilities that constitute a mesoscopic analog of the de~Haas--van~Alphen
effect~\cite{ByersYang1961,AmbegaokarEckern1990}, generated here by spin--orbit coupling rather than an external magnetic field.
In the classical de~Haas--van~Alphen effect, thermodynamic quantities
oscillate as a function of inverse applied field $1/B$; here the
analogous role is played by the coupling~$\xi$, whose variation produces
oscillatory rearrangements of the near-Fermi-edge occupation in
a formally equivalent way.

We further extend the analysis by introducing a phenomenological
interacting model, which allows us to explore the role of collective
effects in the thermodynamic behavior. The resulting framework provides a unified description of spectral, thermodynamic, and mechanical properties in a geometrically constrained quantum system.

Analogous mechanisms may also be realized in cold-atom platforms,
where synthetic spin--orbit coupling and artificial gauge fields have
been demonstrated
experimentally~\cite{LinSpielman2011,DalibardRMP2011,GoldmanRPP2014,GalitskiSpielman2013},
as well as in photonic ring resonators with engineered non-Abelian gauge structures.

This paper is organized as follows.
In Sec.~\ref{Sec:EffectiveModel} we briefly review the effective model
and its spectral properties.
In Sec.~\ref{Sec:Thermo-approach} we develop the thermodynamic formalism
in both canonical and grand-canonical ensembles.
In Sec.~\ref{Sec:InteractingApproach} we introduce the phenomenological
interacting extension.
In Sec.~\ref{Sec:ThermomechanicalResponse} we develop the
thermomechanical framework and derive the corresponding response
coefficients across all three descriptions.
In Sec.~\ref{Sec:Bounds} we connect the effective parameter~$\xi$
to the underlying nonminimal couplings and discuss the order-of-magnitude
experimental bounds from the mesoscopic observables.
Finally, in Sec.~\ref{Sec:Discussion} we discuss the physical
implications of our results and present our conclusions.

\section{Effective model and spectral structure}
\label{Sec:EffectiveModel}

The thermodynamic analysis developed in the subsequent sections is based on the effective mesoscopic model derived in our previous work~\cite{SilvaReisLisboa2025}, where the dynamics of a fermion confined to a one-dimensional quantum ring emerges from nonminimal couplings to antisymmetric tensor fields~\cite{Belich2009,BelichSilva2011,BakkeSilva2012}. For completeness, we briefly summarize the key ingredients of that construction here, emphasizing only the elements that directly affect the statistical and thermodynamic behavior.

After performing the nonrelativistic reduction and imposing confinement to a ring of fixed radius $r_0$, the dynamics is governed by an effective Hamiltonian of the form
\begin{equation}
H
=
\frac{1}{2mr_0^2}\left[
\left(
 i\frac{\partial}{\partial\varphi}
 + \xi\,\sigma_\rho
\right)^2
- \xi^2\right],
\label{eq:Effective-Hamiltonian-summary}
\end{equation}
where $\sigma_\rho$ is the radial Pauli matrix and the dimensionless parameter
\begin{equation}
\xi = m r_0\,\mathcal{F}_{12}
\end{equation}
encodes the strength of the effective spin--orbit interaction induced by the background fields. As shown in the full derivation~\cite{SilvaReisLisboa2025}, this structure can originate either from the tensor $F_{\mu\nu}$ or its dual $\tilde F_{\mu\nu}$, so that both magnetic and electric realizations lead to the same low-energy dynamics.

Equation~\eqref{eq:Effective-Hamiltonian-summary} admits a natural interpretation in terms of a non-Abelian gauge structure: the coupling $\xi\sigma_\rho$ acts as a spin-dependent shift of the angular momentum operator~\cite{Meijer2002,Frustaglia2004}. As a consequence, the orbital motion and the spin degrees of freedom become intrinsically entangled, and the same parameter $\xi$ simultaneously controls spectral shifts, geometric phases~\cite{Berry1984,AharonovCasher1984}, and transport properties.

The corresponding energy spectrum is given by
\begin{equation}
E_{n,s}^{\lambda}
=
\Omega
\left[
\left(
n + \frac{1}{2}
- \frac{\lambda s}{2}\sqrt{1+4\xi^2}
\right)^2
- \xi^2
\right],
\label{eq:Energy-spectrum-summary}
\end{equation}
with $\Omega = (2mr_0^2)^{-1}$, $n\in\mathbb{Z}$, $\lambda=\pm$ labeling the propagation direction, and $s=\pm$ the spin branch.

This spectrum exhibits a characteristic Rashba-like structure~\cite{BychkovRashba1984}, but with a crucial difference: the coupling $\xi$ does not simply split degenerate levels. Instead, it shifts the centers of the angular-momentum parabolas themselves, thereby reorganizing the ordering of low-lying states~\cite{Splettstoesser2003,Molnar2004}. As a result, variations of $\xi$ produce nontrivial rearrangements of the density of states, which are directly probed by thermodynamic quantities.

The corresponding eigenstates display a nontrivial spin texture controlled by a mixing angle $\theta$ defined through
\begin{equation}
\cos\theta = \frac{1}{\sqrt{1+4\xi^2}},
\end{equation}
which determines the degree of spin canting along the ring~\cite{Frustaglia2004,ShengChang2006}. This geometric structure plays a central role in transport and phase effects~\cite{Nagasawa2013,Bergsten2006}, and, as will become clear below, it also governs the thermal response of the system.

From a thermodynamic perspective, the key feature of the model is therefore the $\xi$-dependent deformation of the discrete spectrum in Eq.~\eqref{eq:Energy-spectrum-summary}. Unlike simple rigid shifts of energy levels, this deformation modifies both the spacing and the relative ordering of the states. Consequently, all thermodynamic observables become sensitive to how the effective coupling reshapes the occupation of the low-energy branches.

In what follows, we use this spectrum as the fundamental input for the statistical description. Both canonical and grand-canonical ensembles will be employed to explore complementary physical regimes, allowing us to isolate the role of spectral deformation, finite-size effects, and Fermi statistics in the thermodynamic behavior of the system.

\section{Thermodynamic framework}
\label{Sec:Thermo-approach}

We now develop the thermodynamic description of the system based on the
energy spectrum derived in Sec.~\ref{Sec:EffectiveModel}. Since the
dynamics is confined to a quantum ring, the spectrum is discrete and
labeled by the quantum numbers $(n,s,\lambda)$.

The statistical properties are therefore determined by the partition
function
\begin{equation}
Z = \sum_{n,s,\lambda}
\exp\left(-\beta E_{n,s}^{\lambda}\right),
\label{eq:partition-function}
\end{equation}
where $\beta = 1/(k_B T)$ and the energies $E_{n,s}^{\lambda}$ are given
by Eq.~\eqref{eq:Energy-spectrum-summary}. While the original orbital
quantum number ranges over all integers $n\in\mathbb{Z}$, each distinct
eigenstate is uniquely identified by the triple $(n,s,\lambda)$ with
$n\geq 0$: the spin-dependent shift $c_{s,\lambda}=\frac{1}{2}-\frac{\lambda
s}{2}\sqrt{1+4\xi^2}$ maps the full integer range onto the non-negative
labels without double-counting, as verified explicitly in the companion
work~\cite{SilvaReisLisboa2025}. The quadratic growth of the
spectrum for large $|n|$ ensures the convergence of the partition
function~\cite{PathriaBeal2011,LandauStat}.

\subsection{Role of the effective coupling}

A key feature of the present system is that the parameter $\xi$ does not
produce a simple, rigid splitting of the energy levels. Instead, it
modifies the structure of the spectrum by shifting the centers of the
angular-momentum branches,
\begin{equation}
n \;\rightarrow\;
n + \frac{1}{2}
- \frac{\lambda s}{2}\sqrt{1+4\xi^2}.
\end{equation}

As a consequence, the effective coupling reorganizes the ordering of the
low-energy states~\cite{SilvaReisLisboa2025}. This leads to a nontrivial redistribution of the
occupation probabilities at finite temperature, which directly affects
all thermodynamic observables.

In particular, variations of $\xi$ alter both the spacing and the
relative positioning of the energy levels, effectively deforming the
density of states. This mechanism is fundamentally different from a
simple Zeeman splitting, which is responsible for the distinctive thermal
behavior of the system.

\subsection{Thermodynamic quantities}

Once the partition function is known, the thermodynamic observables
follow from standard relations~\cite{PathriaBeal2011}. The internal energy is given by
\begin{equation}
U = -\frac{\partial}{\partial \beta} \ln Z,
\end{equation}
while the free energy reads
\begin{equation}
F = -\frac{1}{\beta} \ln Z.
\end{equation}

The entropy is obtained as
\begin{equation}
S = -\left( \frac{\partial F}{\partial T} \right),
\end{equation}
and the heat capacity is
\begin{equation}
C = \frac{\partial U}{\partial T}.
\end{equation}

Because the spectrum depends nonlinearly on $\xi$, these quantities
inherit a nontrivial dependence on the effective coupling, providing a
direct probe of the underlying spin--orbit structure.

\subsection{Limiting regimes}

It is instructive to consider two limiting cases.

In the weak-coupling regime $\xi \ll 1$, the spectrum reduces to that of
a free quantum ring with small corrections~\cite{Viefers2004,TanInkson1999}, and the thermodynamic
behavior approaches that of a standard mesoscopic system.

In contrast, in the strong-coupling regime $\xi \gg 1$, the deformation
of the spectrum becomes significant, leading to a redistribution of
low-energy states and a suppression of contributions from higher
branches. This results in characteristic modifications of the heat
capacity and entropy.

These features highlight that the thermodynamics of the system is
governed not only by the presence of spin splitting, but by the
geometric deformation of the spectrum induced by the effective
nonminimal coupling.

\subsection{Thermodynamics}

Having established the single-particle spectrum and eigenstates, we now turn to the thermodynamic properties of a system composed of many non-interacting electrons confined to one-dimensional quantum rings~\cite{Imry2002}. Two complementary ensemble descriptions will be employed:

\begin{itemize}
\item[(i)] the \emph{canonical ensemble}, which describes $N$ independent rings each containing a single electron;  
\item[(ii)] the \emph{grand canonical ensemble}, in which a single ring is populated by $N$ non-interacting fermions obeying the Pauli exclusion principle.
\end{itemize}

These two complementary viewpoints highlight different physical limits.  
In the canonical case, each ring hosts exactly one electron, and therefore, Fermi--Dirac statistics play no role. In the grand canonical formulation, all electrons share the same set of discrete levels, making the exclusion principle essential.
Thermodynamics is especially informative in the present problem because the parameter $\xi$ reshapes the discrete spectrum rather than merely adding a constant offset to all levels. Thermal observables, therefore, probe how the spin-dependent geometric phase reorganizes the density of states and the occupation of the low-lying branches.
\vspace{0.2cm}

\noindent\textbf{Canonical ensemble.}  
The canonical partition function for $N$ non-interacting rings factorizes into $N$ copies of the single-particle partition function,
\begin{equation}
\mathcal{Z}
 = \sum_{\{n_{j}\}}
   \exp\left(-\beta E_{\{n_{j}\}}\right)
 = \mathcal{Z}_{1}^{N},
\label{eq:Partition-function}
\end{equation}
The factorization $\mathcal Z=\mathcal Z_1^N$ expresses the physical assumption that different rings are independent in the canonical ensemble. As a result, all equilibrium properties of the whole array are already encoded in the one-electron spectrum of a single ring.
where
\begin{equation}
\mathcal{Z}_{1}
 = \sum_{n=0}^{\infty}
   \sum_{s=\{\pm\}}
   \sum_{\lambda=\{\pm\}}
   \exp\left( -\beta E_{n,s}^{\lambda} \right).
\label{eq:Single-Partition-function}
\end{equation}
The relevant thermodynamic quantity is the \emph{Helmholtz free energy per particle},
\begin{equation}
\mathcal{F}
 = -\frac{1}{\beta}\lim_{N\to\infty}\frac{1}{N}\ln\mathcal{Z}.
\end{equation}
From $\mathcal{F}$ we may compute the internal energy, entropy, heat capacity, as well as the analogs of magnetization and susceptibility induced by the spin--orbit parameter $\xi$. Throughout this section, we employ natural units, so $k_{B}=1$.

\vspace{0.2cm}

\noindent\textbf{Grand canonical ensemble.}  
In the grand canonical formulation, the grand partition function is
\begin{equation}
\Xi
 = \sum_{N=0}^{\infty}
   e^{\beta\mu N}\,
   \mathcal{Z}\big[N_{n,s}^{\lambda}\big],
\label{eq:GrandPartition-function}
\end{equation}
where the occupation numbers satisfy $N_{n,s}^{\lambda} \in \{0,1\}$ for fermions~\cite{PathriaBeal2011}.  
The many-particle energy is
\begin{equation*}
E\{N_{n,s}^{\lambda}\}
 = \sum_{n,s,\lambda}
   N_{n,s}^{\lambda}\,E_{n,s}^{\lambda},
\qquad
\sum_{n,s,\lambda} N_{n,s}^{\lambda}=N.
\end{equation*}

The canonical partition function for fixed occupations becomes
\begin{equation}
\mathcal{Z}[N_{n}]
 = \sum_{\{N_{n,s}^{\lambda}\}}
   \exp\!\left[
     -\beta
     \sum_{n,s,\lambda}
     N_{n,s}^{\lambda}E_{n,s}^{\lambda}
   \right].
\end{equation}

Inserting this expression into Eq.~\eqref{eq:GrandPartition-function} yields
\begin{align}
\Xi
 &= \prod_{n,s,\lambda}
   \left\{
     \sum_{N_{n,s}^{\lambda}=0}^{1}
     \exp\!\left[
       -\beta N_{n,s}^{\lambda}(E_{n,s}^{\lambda}-\mu)
     \right]
   \right\}
\\
 &= \prod_{n,s,\lambda}
   \left[
     1 + \exp\!\left( -\beta (E_{n,s}^{\lambda}-\mu) \right)
   \right].
\end{align}

The corresponding grand potential is
\begin{align}
\Phi
 &= -\frac{1}{\beta}\ln\Xi \notag \\
 &= -\frac{1}{\beta}
   \sum_{n=0}^{\infty}
   \sum_{s=\{\pm\}}
   \sum_{\lambda=\{\pm\}}
   \ln\!\left[
     1
     + \exp\left(
         -\beta (E_{n,s}^{\lambda}-\mu)
       \right)
   \right].
\label{eq:Grand-potential}
\end{align}
In the grand canonical description, the same split spectrum is interrogated from a different viewpoint: many fermions compete for the available branches, and the chemical potential selects which part of the spectrum is most relevant. This makes the grand potential especially sensitive to small shifts near the Fermi edge, which is why oscillatory low-temperature responses are amplified in this ensemble~\cite{CheungGefenRiedel1988,AmbegaokarEckern1990}.
From $\Phi$ one obtains the thermodynamic observables, including the mean particle number, internal energy, entropy, heat capacity, and the analog of magnetization and susceptibility.

\vspace{0.3cm}

\noindent\textbf{Summation over energy levels.}  
Both ensemble descriptions require the evaluation of infinite sums over the discrete quantum number $n$. These sums cannot be performed in closed form due to the non-quadratic structure of the spectrum. To preserve finite-size effects, which are crucial in one-dimensional rings~\cite{Viefers2004}, we do not take the thermodynamic limit $n\to\infty$ in the sum. Instead, we use the Euler--Maclaurin summation formula,
\begin{align}
\sum_{n=0}^{\infty} F(n)
 &= \int_{0}^{\infty}F(n)\, dn
    + \frac{1}{2}F(0)
    - \frac{1}{2!}B_{2}F'(0)
    + \cdots ,
\end{align}
where $B_{m}$ are the Bernoulli numbers ($B_{2}=1/6$, $B_{4}=-1/30$, $\ldots$). Truncating this expansion after a few terms provides an accurate approximation, as confirmed numerically.
The usefulness of the Euler--Maclaurin method here is that it keeps the mesoscopic discreteness of the ring while still yielding compact analytical control. The integral term captures the smooth global trend, whereas the boundary and derivative corrections retain the finite-size information responsible for the low-temperature structure of the observables.

The next subsections explicitly implement this method in both the canonical and grand canonical ensembles, yielding analytical expressions for the partition functions and enabling direct computation of thermodynamic quantities.

\subsection{Canonical approach}

We now implement the Euler--Maclaurin summation formula to evaluate the single-particle canonical partition function,
\begin{equation}
\mathcal{Z}_{1}
 = \sum_{n=0}^{\infty}
   \left[
     e^{-\beta E_{n,+}^{+}}
     + e^{-\beta E_{n,-}^{+}}
     + e^{-\beta E_{n,+}^{-}}
     + e^{-\beta E_{n,-}^{-}}
   \right],
\end{equation}
which allows us to identify the summand as
\begin{align}
F(n)
 &= e^{-\beta E_{n,+}^{+}}
   + e^{-\beta E_{n,-}^{+}}
   + e^{-\beta E_{n,+}^{-}}
   + e^{-\beta E_{n,-}^{-}}.
\end{align}
The four contributions correspond to the two propagation directions and the two spin branches. Because $\xi$ shifts these branches unequally, the canonical partition function is governed not by a single excitation gap but by a family of nearby energy scales that become thermally active at different temperatures.
\bigskip

\noindent\textbf{Euler--Maclaurin expansion.}  
Introducing the substitution $u = n + c_{s,\lambda}$ with
$c_{s,\lambda}=\frac{1}{2}-\frac{\lambda s}{2}\sqrt{1+4\xi^2}$, each branch integral takes the form
$e^{\beta\Omega\xi^{2}}\int_{c_{s,\lambda}}^{\infty}e^{-\beta\Omega u^{2}}du$,
since the $-\xi^{2}$ term in the spectrum supplies the common prefactor
$e^{\beta\Omega\xi^{2}}$ after the variable shift.
The leading contribution arises from the integral
\begin{align}
&\int_{0}^{\infty} F(n)\, dn
 = \sqrt{\frac{\pi}{\Omega\beta}}
    \exp\!\left( \beta\Omega\xi^{2} \right)
\notag\\
 &\quad\times
    \left\{
      2
      + \operatorname{erf}\!\left( \sqrt{\Omega\beta}\,\Lambda_{-}\right)
      - \operatorname{erf}\!\left( \sqrt{\Omega\beta}\,\Lambda_{+}\right)
    \right\},
\label{eq:integral-term}
\end{align}
where
\begin{equation}
\Lambda_{\pm}
 = \frac{1}{2}
   \left( \sqrt{1+4\xi^{2}} \pm 1 \right),
\end{equation}
and
\begin{equation}
\operatorname{erf}(z)
 = \frac{2}{\sqrt{\pi}}\!\int_{0}^{z} e^{-t^{2}}\, dt.
\end{equation}

The error functions appearing in the integral term are a direct fingerprint of the shifted parabolic spectrum. The parameters $\Lambda_{\pm}$ act as effective thresholds that delimit how the two split branches enter the thermal sum. The next two Euler--Maclaurin terms are

\begin{align}
\frac{1}{2}F(0)
 &= e^{\Omega\beta\Lambda_{-}}
  + e^{-\Omega\beta\Lambda_{+}}, \\
-\frac{1}{2!}B_{2}F'(0)
 &= \frac{\Omega\beta}{6}
    \left[
       2\Lambda_{+}e^{-\Omega\beta\Lambda_{+}}
     - 2\Lambda_{-}e^{\Omega\beta\Lambda_{-}}
    \right].
\end{align}

Combining these contributions yields
\begin{align}
\mathcal{Z}_{1}
 &= \sqrt{\frac{\pi}{\Omega\beta}}\exp\!\left( \beta\Omega\xi^{2} \right)\times\notag\\
 &   \times\left\{
      2
      + \operatorname{erf}\!\left( \sqrt{\Omega\beta}\,\Lambda_{-}\right)
      - \operatorname{erf}\!\left( \sqrt{\Omega\beta}\,\Lambda_{+}\right)
    \right\}+\notag\\
 &\quad
  + \left( 1 - \frac{\Omega\beta}{3}\Lambda_{-} \right)e^{\Omega\beta\Lambda_{-}}
  + \left( 1 + \frac{\Omega\beta}{3}\Lambda_{+} \right)e^{-\Omega\beta\Lambda_{+}}.
\label{eq:Z1-full}
\end{align}

\begin{figure}[h]
\centering
\includegraphics[width=8cm,height=5cm]{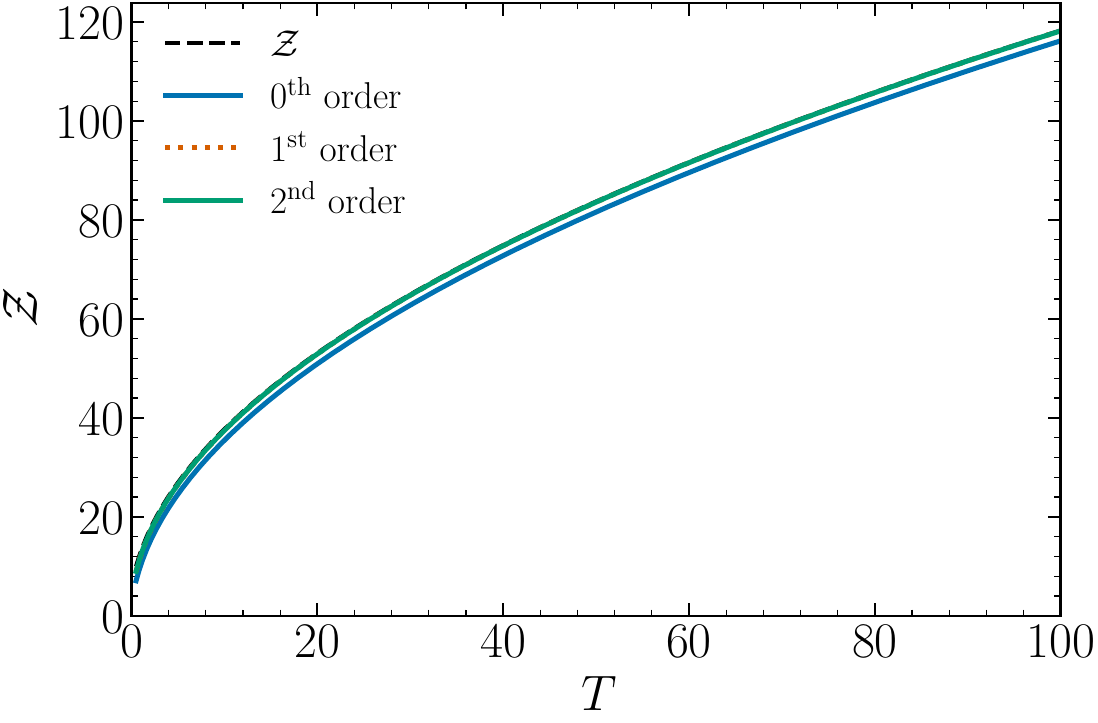}
\caption{Single-particle canonical partition function $\mathcal Z_{1}$ as a function of temperature. The dashed curve represents the direct numerical evaluation obtained by summing over a large number of energy levels, while the solid curves show successive truncations of the Euler--Maclaurin expansion at zeroth, first, and second order. Already at first order, the analytic approximation is virtually indistinguishable from the exact result over the entire temperature range, confirming that the method captures both the low-temperature discreteness and the high-temperature continuum limit with remarkable precision.}
\label{Fig:NumComp}
\end{figure}

This truncated expression already provides excellent accuracy, as shown in Fig.~\ref{Fig:NumComp}. Its physical interpretation is straightforward: at low temperature, the exponential pieces isolate the lowest split branches, and the dependence on $\xi$ is dominated by the ground-state displacement, whereas at higher temperature, the square-root prefactor and the error functions describe the gradual activation of a wider set of angular-momentum states.

The rapid convergence of the Euler--Maclaurin series, visible in Fig.~\ref{Fig:NumComp}, has two important consequences. First, it validates the use of the analytical expressions in all subsequent thermodynamic calculations. Second, it demonstrates that the leading integral term already captures the dominant $\sqrt{T}$ growth of $\mathcal Z_{1}$, which originates from the quadratic dispersion of the angular-momentum branches, while the boundary and derivative corrections efficiently encode the mesoscopic finite-size effects that distinguish this system from a true continuum.
\bigskip

\begin{figure}[h]
\centering
\includegraphics[width=8cm,height=5cm]{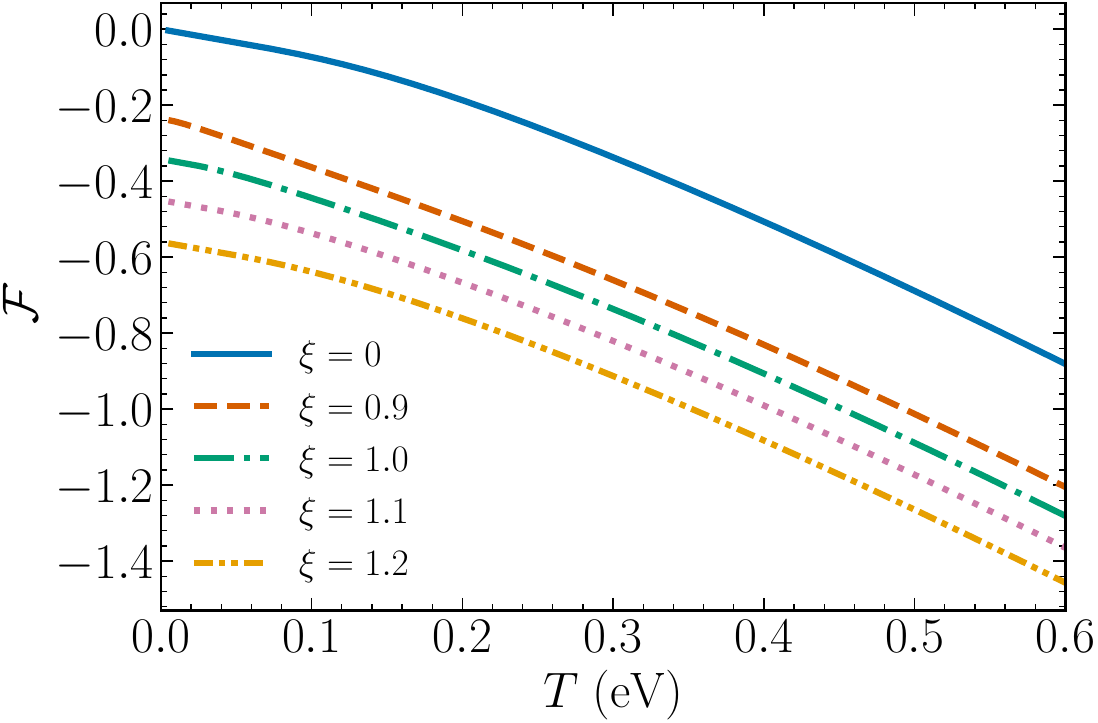}
\caption{Helmholtz free energy per quantum ring as a function of temperature for representative values of the effective coupling $\xi$. The monotonic decrease reflects the growth of thermally accessible microstates as temperature increases. Larger values of $\xi$ lower the free energy throughout the plotted range, with the most pronounced separation occurring at low temperature, where only the first few mesoscopic branches contribute.}
\label{fig:FPlot}
\end{figure}

\noindent\textbf{Helmholtz free energy and thermodynamic functions.}  
With $\mathcal{Z}_{1}$ known, the Helmholtz free energy per ring is
\begin{equation*}
\mathcal{F}
 = -\frac{1}{\beta}\ln\mathcal{Z}_{1}.
\end{equation*}
The behavior of $\mathcal{F}$ as a function of temperature is shown in Fig.~\ref{fig:FPlot}.

The behavior of $\mathcal{F}(T)$ displayed in Fig.~\ref{fig:FPlot} can be understood as follows. At $T=0$, the free energy equals the ground-state energy, which the Rashba-like splitting shifts downward for finite $\xi$. As the temperature rises, additional angular-momentum states become thermally populated and the entropy contribution $-TS$ drives $\mathcal{F}$ to increasingly negative values. The fact that curves for different $\xi$ converge at high temperature indicates that the detailed low-energy splitting becomes less relevant once many levels participate in the thermal average, and the system approaches the behavior of a quasi-continuous spectrum.

The remaining thermodynamic functions---entropy, internal energy, and heat capacity---follow from standard relations:
\begin{align*}
S = -\left( \frac{\partial \mathcal{F}}{\partial T} \right)_{V},\quad
U = \mathcal{F} + TS,\quad
C = \left( \frac{\partial U}{\partial T} \right)_{V}.
\end{align*}
Although their analytical forms are lengthy, their behavior is shown below. These standard derivatives acquire a particularly rich structure here because the underlying spectrum is not a simple rigid ladder: every derivative with respect to temperature probes how the coupling-dependent level shifts modify the thermal occupation of the split branches.
\begin{figure}[h!]
\centering
\includegraphics[width=8cm,height=5cm]{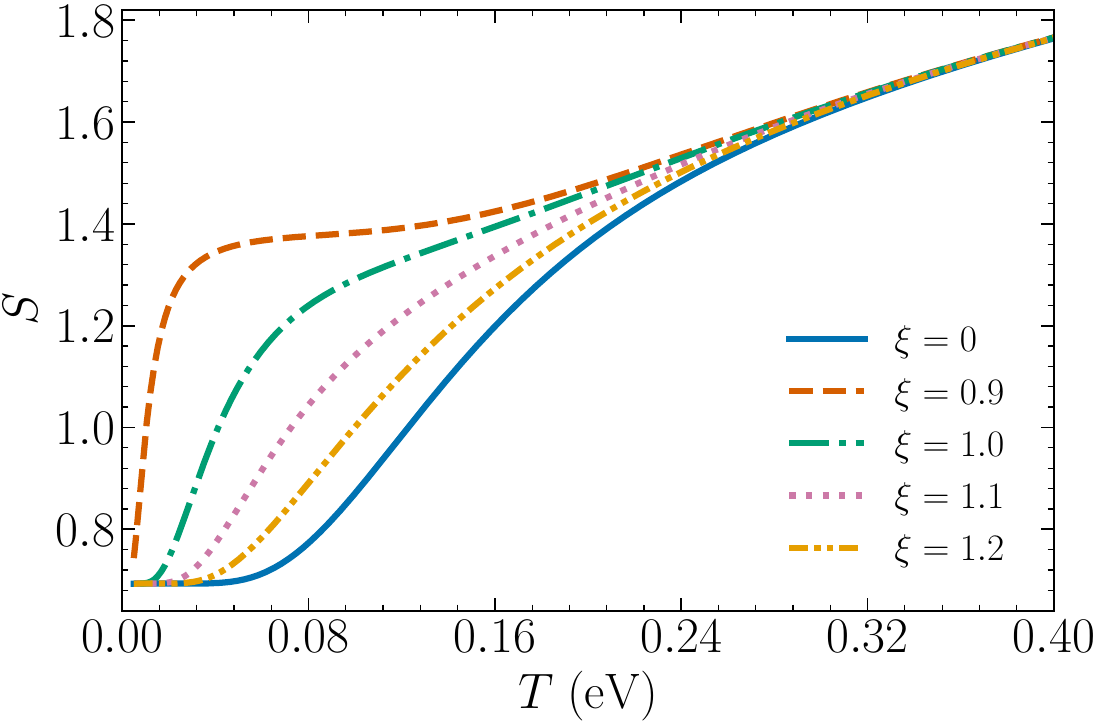}
\caption{Entropy per quantum ring as a function of temperature for representative values of $\xi$. Finite values of the effective coupling unlock the first excited states at lower temperatures, producing an earlier and steeper onset of entropy compared with the uncoupled case.}
\label{fig:SPlot}
\end{figure}

The entropy curves in Fig.~\ref{fig:SPlot} exhibit a characteristic sigmoid-like growth whose inflection point shifts to lower temperatures as $\xi$ increases. This can be traced to the reorganization of the near-ground-state manifold: larger $\xi$ brings the lowest split levels closer together, so that thermal fluctuations can populate them at a reduced energy cost. At higher temperatures all curves converge, reflecting the universal high-energy behavior of the quadratic dispersion once the mesoscopic details are thermally averaged out. The low-temperature value $S\to\ln 2$ for $\xi=0$ corresponds to the residual degeneracy of the two spin-propagation channels that share the ground-state energy.
\begin{figure}[h!]
\centering
\includegraphics[width=8cm,height=5cm]{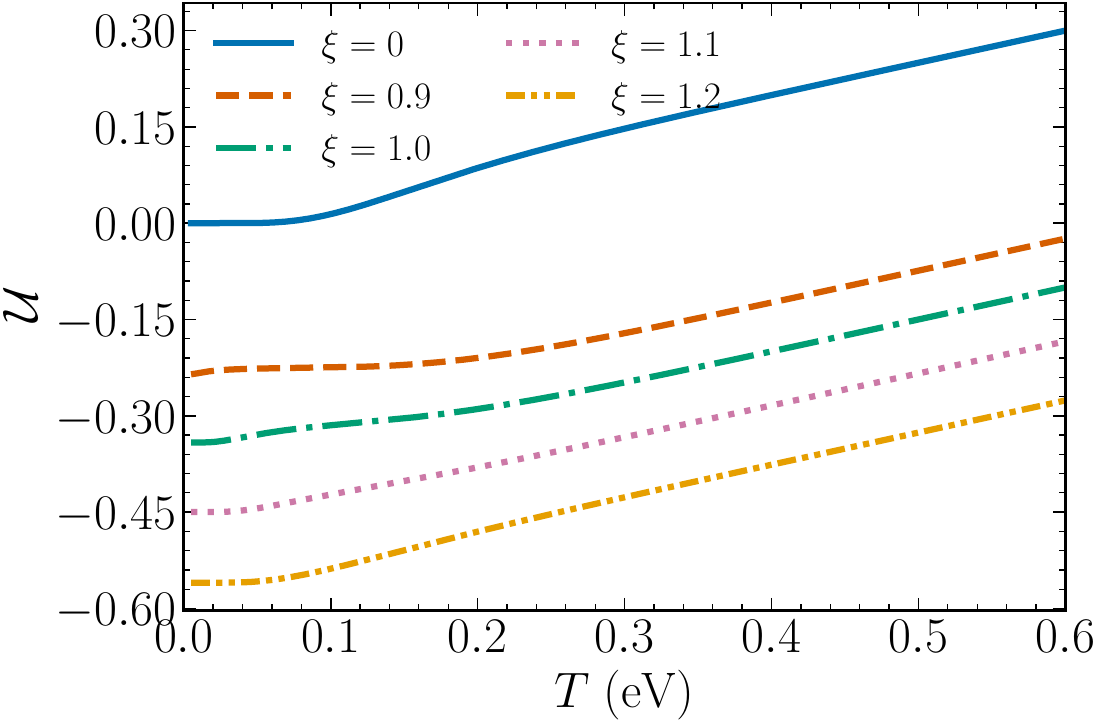}
\caption{Internal energy per quantum ring versus temperature for several representative values of $\xi$. The vertical offset between curves originates from the $\xi$-dependent shift of the ground-state energy, while the common positive slope at higher temperatures reflects the progressive thermal occupation of excited angular-momentum states.}
\label{fig:UPlot}
\end{figure}

As seen in Fig.~\ref{fig:UPlot}, the internal energy at low temperature is essentially equal to the coupling-shifted ground-state energy; consequently, larger $\xi$ displaces the curve downward. With increasing temperature, the curves show a nearly linear increase, characteristic of a dense ladder of states being progressively populated. The transition region between the plateau and the linear regime coincides with the temperature at which the first excited branch becomes thermally active, and its location is directly controlled by $\xi$. This interplay between spectral geometry and thermal activation makes $U(T)$ a particularly transparent diagnostic of the effective spin--orbit interaction.
\begin{figure}[h!]
\centering
\includegraphics[width=8cm,height=5cm]{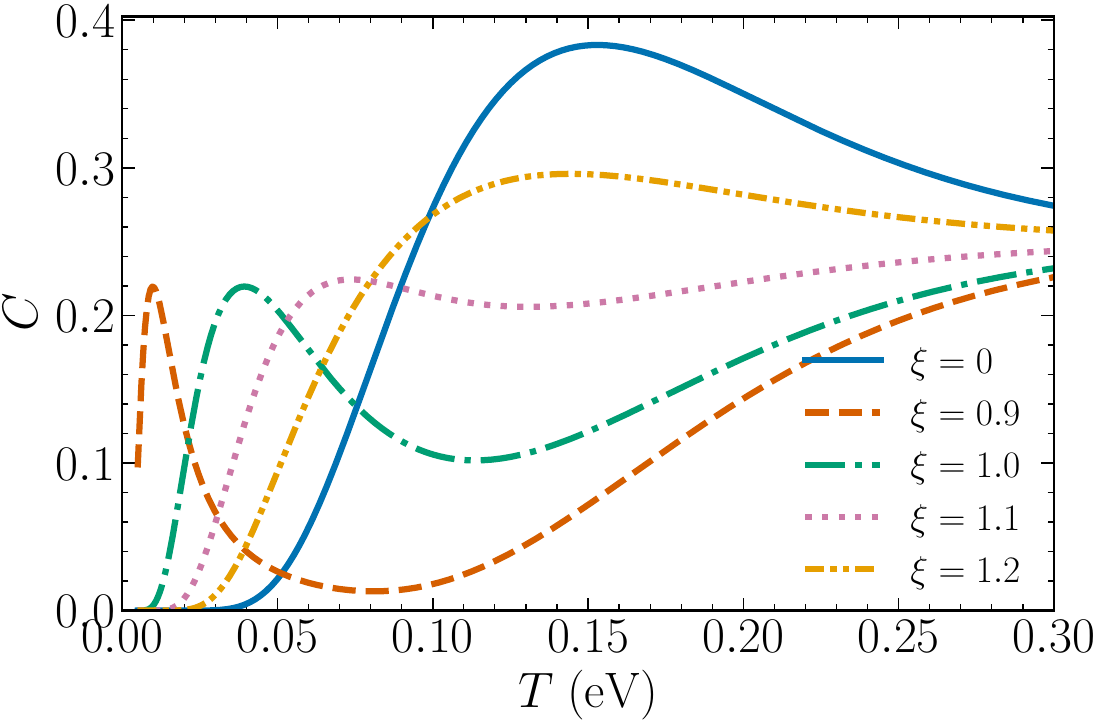}
\caption{Heat capacity per quantum ring as a function of temperature for representative couplings. The Schottky-like peaks identify the temperature scales at which the thermal population of the split mesoscopic levels is redistributed most rapidly. For finite $\xi$, a secondary low-temperature peak emerges from the activation of states whose energy is set by the spin--orbit gap.}
\label{fig:CPlot}
\end{figure}

The heat capacity, shown in Fig.~\ref{fig:CPlot}, provides the clearest thermodynamic fingerprint of the spectral reconstruction induced by the background field. For $\xi=0$, a single broad Schottky-like peak is observed, centered at a temperature comparable to the first excitation gap $\Omega$. As $\xi$ increases, this peak splits and moves to lower temperatures, signaling the emergence of an additional low-energy scale set by the Rashba-like splitting. The peak height is determined by the number of nearly degenerate states that participate in the thermal redistribution, while its width reflects the dispersion of their energies. In this sense, $C(T)$ acts as a spectroscopic probe of the low-energy branch structure, providing information that is complementary to transport measurements~\cite{TanInkson1999,Lorke2000}.
\bigskip

\noindent\textbf{Magnetization and susceptibility.}  
The analog of the magnetization is obtained by differentiating the free energy with respect to the Rashba-like coupling parameter $\xi$:
\begin{equation}
M = -\left( \frac{\partial \mathcal{F}}{\partial \xi} \right)_{T}.
\end{equation}
The corresponding susceptibility is
\begin{equation}
\chi = \left( \frac{\partial M}{\partial \xi} \right)_{T}.
\end{equation}
Both quantities exhibit distinctive low-temperature features characteristic of one-dimensional rings with spin--orbit coupling~\cite{Frustaglia2004,Splettstoesser2003}. Since both observables are derivatives with respect to $\xi$, they directly measure how strongly the spectrum reacts to changes in the effective Rashba parameter and are therefore especially sensitive to near-degeneracies and level rearrangements.
\begin{figure}[h!]
\centering
\includegraphics[width=8cm,height=5cm]{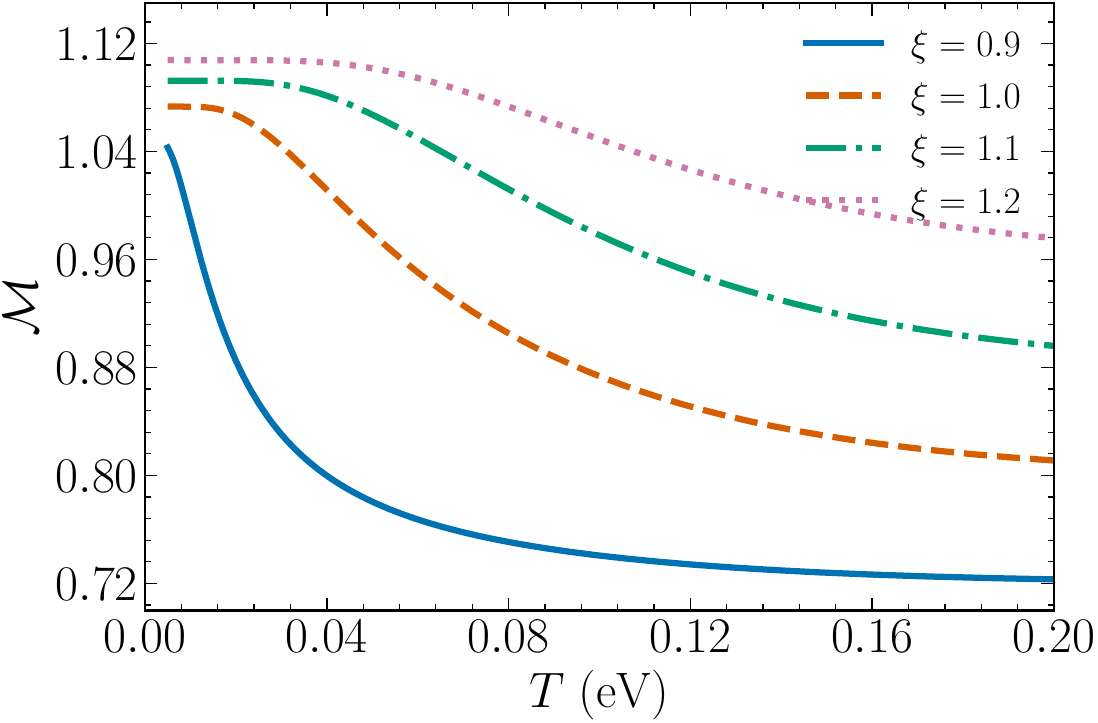}
\caption{Canonical magnetization analog $M=-\partial\mathcal F/\partial\xi$ as a function of temperature for representative couplings. The quantity measures the thermodynamic response of the free energy to variations of the effective Rashba parameter rather than to a physical magnetic flux.}
\label{fig:MPlot}
\end{figure}

Figure~\ref{fig:MPlot} shows that the magnetization analog is largest at low temperature, where the free energy is dominated by a small number of split levels and is therefore most sensitive to spectral displacements induced by $\xi$. As the temperature increases, thermal averaging progressively washes out the coupling sensitivity, and the curves approach a common plateau. The smooth monotonic evolution confirms that the canonical ensemble does not exhibit the sharp filling-driven features that are expected in the grand canonical case.
\begin{figure}[h!]
\centering
\includegraphics[width=8cm,height=5cm]{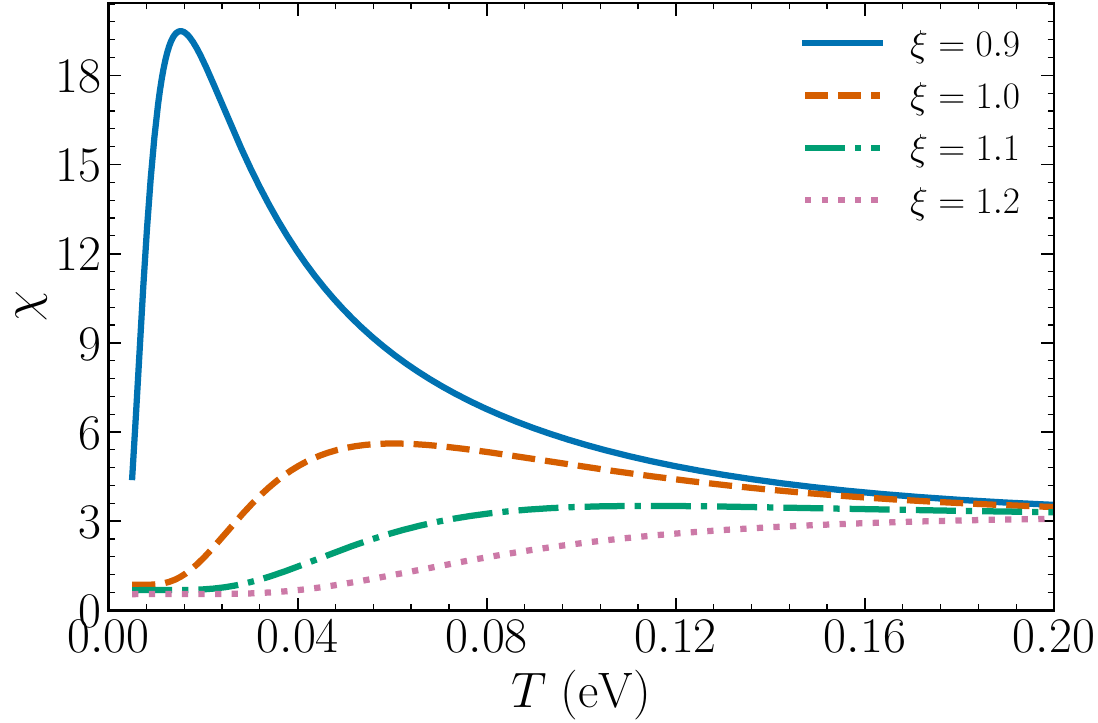}
\caption{Canonical susceptibility analogue $\chi=\partial M/\partial\xi$ as a function of temperature for representative couplings. Pronounced low-temperature peaks indicate a regime of enhanced sensitivity of the mesoscopic spectrum to small variations of the effective spin--orbit strength.}
\label{fig:XPlot}
\end{figure}

The susceptibility displayed in Fig.~\ref{fig:XPlot} exhibits a sharp peak at low temperature whose position and amplitude depend on $\xi$. This enhanced response arises because, at low temperatures, only a few closely spaced levels contribute to the partition function, so that a small change in the coupling can appreciably reshape the Boltzmann weights. As the temperature rises and the partition function becomes a broad thermal average over many levels, the susceptibility decays and the sensitivity to $\xi$ is lost. The peak, therefore, serves as a direct indicator of the temperature regime in which the mesoscopic spectral structure is most relevant.
\bigskip

\begin{figure}[h!]
\centering
\includegraphics[width=8cm,height=5cm]{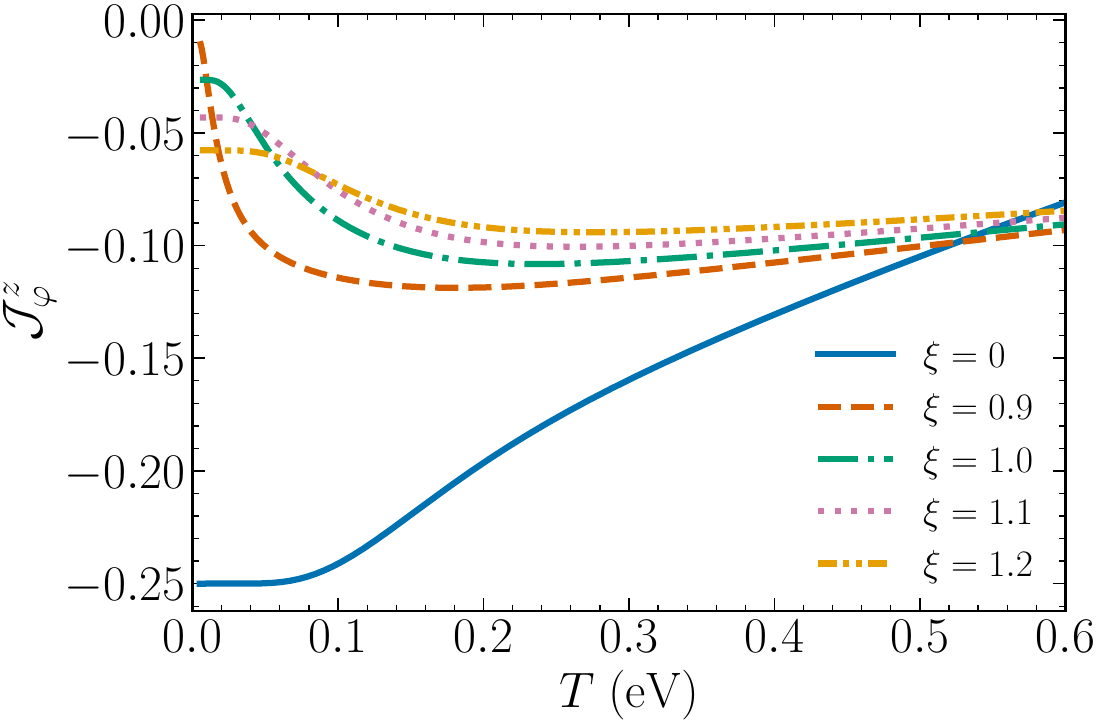}
\caption{Canonical thermal expectation value of the $z$ component of the persistent spin current for representative values of $\xi$. The competition between geometric spin canting and thermal broadening governs the evolution from the strongly coupling-dependent low-temperature regime to the smoother high-temperature behavior.}
\label{fig:JThermal}
\end{figure}

\noindent\textbf{Thermal dependence of the spin current.}  
Using the canonical ensemble, the thermal expectation value of the $z$-component of the persistent spin current~\cite{ButtikerImryLandauer1983,Splettstoesser2003,FoersterBuettiker2008} is
\begin{subequations}
\begin{align*}
\langle \mathcal{J}_{\varphi}^{z} \rangle
 &= \frac{1}{\mathcal{Z}_{1}}
    \sum_{n=0}^{\infty}
    \sum_{s,\lambda=\pm}
    \mathcal{J}_{\varphi}^{z}
    e^{-\beta E_{n,s}^{\lambda}}
\\
 &= \frac{1}{\mathcal{Z}_{1}}
    \sum_{n=0}^{\infty}
    \sum_{s,\lambda=\pm}
    \frac{1}{4mr_{0}}
    \left(2n\cos\theta - 1\right)
    e^{-\beta E_{n,s}^{\lambda}}
\\
 &= \frac{2\cos\theta}{4mr_{0}}
    \frac{1}{\mathcal{Z}_{1}}
    \sum_{n=0}^{\infty}
    \sum_{s,\lambda=\pm}
    n\,e^{-\beta E_{n,s}^{\lambda}}
    - \frac{1}{4mr_{0}}.
\end{align*}
\end{subequations}
Its thermal behavior is shown in Fig.~\ref{fig:JThermal}.

The spin current displayed in Fig.~\ref{fig:JThermal} reveals a clear separation between two regimes. At low temperature the expectation value depends strongly on $\xi$ because the occupation is concentrated on the lowest branches, whose spin polarization is directly controlled by the mixing angle $\theta$~\cite{Frustaglia2004,Nagasawa2013}. For $\xi=0$ the spin is fully aligned along the $z$~axis and the current retains a comparatively large magnitude, while finite coupling rotates the spin texture away from $z$ and reduces the net polarized transport. As the temperature increases, higher angular-momentum states---with oscillating spin projections---are progressively populated, and the thermal average tends toward a common, smoother curve that is less sensitive to the details of the low-energy splitting.

\subsection{Grand canonical approach}

We now perform the thermodynamic analysis using the grand canonical ensemble. In this case, the relevant sums are more involved due to the logarithmic structure of the integrand in the grand potential,
\begin{equation}
\Phi
 = -\frac{1}{\beta}
   \sum_{n=0}^{\infty}
   \sum_{s,\lambda=\pm}
   \ln \left[
        1 + \exp\left(
              -\beta (E_{n,s}^{\lambda}-\mu)
            \right)
      \right].
\end{equation}
Because the logarithm weighs the occupation of every branch through the Fermi--Dirac distribution, this sum is highly sensitive to spectral shifts near the chemical potential. Small changes in $\xi$ can therefore produce sizable changes in the grand-canonical observables by modifying which split levels lie closest to the Fermi edge.

Accordingly, we introduce the function
\begin{align}
F(n)
 &= \ln\!\left[
        1 + e^{-\beta(E_{n,+}^{+}-\mu)}
     \right]
  + \ln\!\left[
        1 + e^{-\beta(E_{n,-}^{+}-\mu)}
     \right]
\notag\\
 &\quad
  + \ln\!\left[
        1 + e^{-\beta(E_{n,+}^{-}-\mu)}
     \right]
  + \ln\!\left[
        1 + e^{-\beta(E_{n,-}^{-}-\mu)}
     \right].
\end{align}

\bigskip

\noindent\textbf{Integral term of the Euler--Maclaurin formula.}  
Because the spectrum is quadratic in $n$ up to a shift, the integral
\begin{equation}
\int_{0}^{\infty} F(n)\, dn
\end{equation}
may be rewritten in a more convenient form by integrating by parts. After doing so, one arrives at
\begin{align}
\int_{0}^{\infty} F(n)\, dn
 &= \int_{\Lambda_{+}}^{\infty}
    \frac{2\beta u(u-\Lambda_{+})}
         {\Delta^{-1} e^{\beta u^{2}} + 1}\, du
\notag\\
 &\quad
  + \int_{\Lambda_{-}}^{\infty}
    \frac{2\beta u(u-\Lambda_{-})}
         {\Delta^{-1} e^{\beta u^{2}} + 1}\, du,
\label{eq:Fn-GrandC}
\end{align}
where
\begin{equation}
\Delta = \exp\big[ \beta(\mu + \Omega \xi^{2}) \big].
\end{equation}
The lower limits $\Lambda_{\pm}$ show again that the coupling alters the effective thresholds from which the two split branches begin to contribute. In the grand canonical ensemble, these thresholds compete directly with the chemical potential and therefore strongly influence the filling pattern.

The Fermi--Dirac kernel may be expanded as
\begin{equation}
\frac{1}{\Delta^{-1} e^{\beta u^{2}} + 1}
 = \sum_{k=1}^{\infty}
   (-1)^{k-1}\,
   \Delta^{k}\,
   e^{-k \beta u^{2}}.
\end{equation}
Inserting this series into Eq.~\eqref{eq:Fn-GrandC} yields
\begin{align}
&\int_{0}^{\infty} F(n)\, dn
 = 2\beta
    \sum_{k=1}^{\infty}
    (-1)^{k-1}\Delta^{k}\notag\\ &\times
    \int_{\Lambda_{+}}^{\infty}
       u(u-\Lambda_{+}) e^{-k\beta u^{2}}\, du
\notag\\
 &\quad
  +2\beta
    \sum_{k=1}^{\infty}
    (-1)^{k-1}\Delta^{k}
    \int_{\Lambda_{-}}^{\infty}
       u(u-\Lambda_{-}) e^{-k\beta u^{2}}\, du.
\end{align}

Both integrals can be evaluated analytically, resulting in
\begin{align}
&\int_{0}^{\infty} F(n)\, dn
 = \sum_{k=1}^{\infty}
    (-1)^{k-1}
    \Delta^{k}
    \frac{1}{2}
    \sqrt{\frac{\pi}{k^{3}\beta}}
\notag\\
 &\quad\times
    \left\{
      2
      - \operatorname{erf}\!\left( \sqrt{k\beta}\,\Lambda_{+} \right)
      - \operatorname{erf}\!\left( \sqrt{k\beta}\,\Lambda_{-} \right)
    \right\}.
\end{align}
\bigskip

\noindent\textbf{Remaining Euler--Maclaurin terms.}  
The boundary term reads
\begin{align}
&\frac{1}{2}F(0)
 = \ln\!\left[
       1 + e^{-\beta(\Omega\Lambda_{+}-\mu)}
    \right]
 + \ln\!\left[
       1 + e^{\beta(\Omega\Lambda_{-} + \mu)}
    \right],
\end{align}
while the first derivative term gives
\begin{align}
-\frac{1}{2!}B_{2}F'(0)
 &= \frac{\Omega\beta}{3\Upsilon}
    \left[
      \Lambda_{+}
      + e^{\beta\mu} e^{\beta\Omega\Lambda_{-}}
      - \Lambda_{-} e^{\beta\Omega(\Lambda_{+}+\Lambda_{-})}
    \right],
\end{align}
with
\begin{equation}
\Upsilon
 = \left( 1 + e^{-\beta\mu} e^{\beta\Omega\Lambda_{+}} \right)
   \left( 1 + e^{\beta\mu} e^{\beta\Omega\Lambda_{-}} \right).
\end{equation}

Putting all contributions together yields the grand potential
\begin{align}
\Phi
 &= -\sum_{k=1}^{\infty}
      (-1)^{k-1}\Delta^{k}
      \frac{1}{2}
      \sqrt{\frac{\pi}{k^{3}\beta^{3}}}
\notag\\
 &\quad\times
      \left\{
        2
        - \operatorname{erf}\!\left(\sqrt{k\beta}\,\Lambda_{+}\right)
        - \operatorname{erf}\!\left(\sqrt{k\beta}\,\Lambda_{-}\right)
      \right\}
\notag\\
 &\quad
   - \frac{1}{\beta}
     \ln\!\left[
       \frac{
         1 + e^{-\beta(\Omega\Lambda_{+}-\mu)}
       }{
         1 + e^{\beta(\Omega\Lambda_{-}+\mu)}
       }
     \right]
\notag\\
 &\quad
   - \frac{\Omega}{3\Upsilon}
     \left[
       \Lambda_{+}
       + e^{\beta\mu}e^{\beta\Omega\Lambda_{-}}
       - \Lambda_{-} e^{\beta\Omega(\Lambda_{+}+\Lambda_{-})}
     \right].
\end{align}

Although a closed-form expression is not possible due to the infinite series, the representation above is efficient numerically and reveals the key physical behavior of the system. The alternating series is the analytic imprint of Fermi statistics, while the error functions retain the structure of the shifted parabolic spectrum. Keeping both ingredients is essential: the first captures the exclusion principle, and the second preserves the mesoscopic finite-size corrections that would be lost in a continuum approximation.
\bigskip

\noindent\textbf{Thermodynamic functions.}  
All grand canonical thermodynamic quantities follow from derivatives of $\Phi(T,\mu)$:

\begin{align*}
N &= -\left( \frac{\partial\Phi}{\partial\mu} \right)_{T},
&
U &= \Phi + TS + \mu N,
\\
S &= -\left( \frac{\partial\Phi}{\partial T} \right)_{\mu},
&
C &= \left( \frac{\partial U}{\partial T} \right)_{\mu}.
\end{align*}

The numerical behavior of entropy, internal energy, heat capacity, magnetization, and susceptibility is illustrated in the figures below.

\begin{figure}[h!]
\centering
\includegraphics[width=8cm,height=5cm]{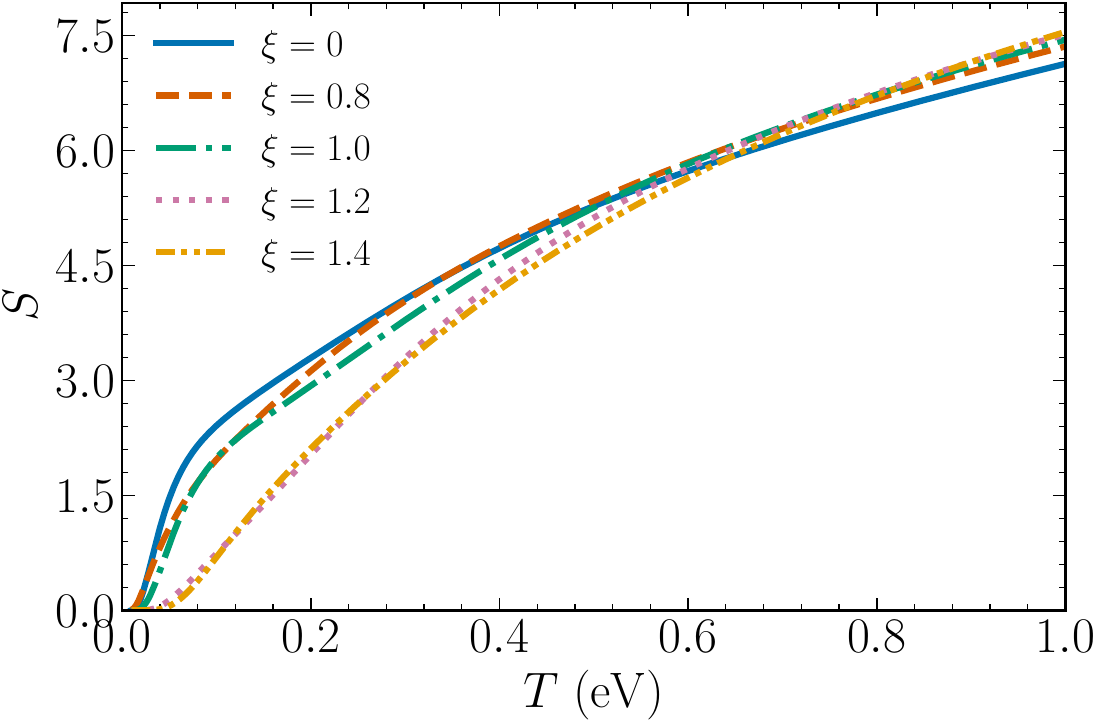}
\caption{Grand-canonical entropy as a function of temperature for representative values of $\xi$ at fixed chemical potential. The ordering of the curves at low temperature is determined by the number of states below the Fermi edge for each coupling, while the high-temperature convergence reflects the universal growth of accessible configurations.}
\label{fig:SPlotC2}
\end{figure}

The grand-canonical entropy in Fig.~\ref{fig:SPlotC2} is a sensitive probe of the low-temperature filling pattern. For coupling values that place a near-degenerate pair of levels at the chemical potential, a finite residual entropy persists at low~$T$, signaling incomplete Fermi-occupation ordering. Other couplings open a gap at the Fermi edge and drive the entropy toward zero. As the temperature rises, these differences are progressively washed out because the number of thermally accessible states grows faster than the splitting can resolve, and all curves converge onto a common envelope governed by the effective density of states.
\begin{figure}[h!]
\centering
\includegraphics[width=8cm,height=5cm]{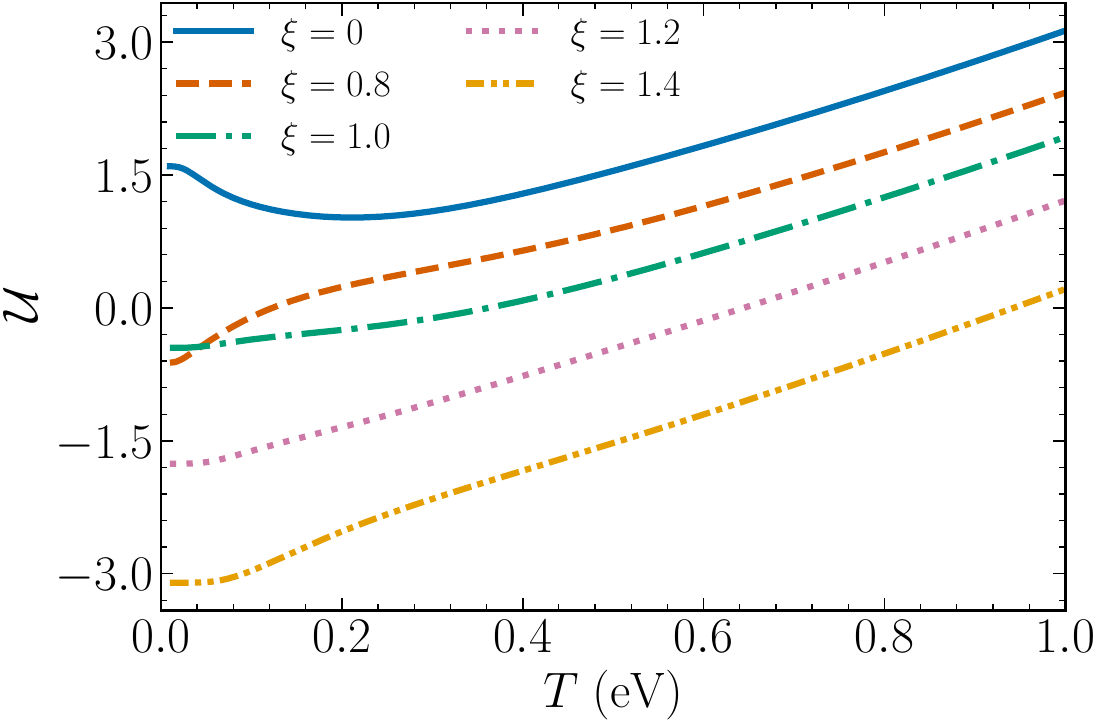}
\caption{Grand-canonical internal energy as a function of temperature for representative values of $\xi$. The large vertical separation between the curves arises because the coupling simultaneously shifts the energy levels and modifies which branches are occupied at the given chemical potential.}
\label{fig:UPlotC2}
\end{figure}

The grand-canonical internal energy, shown in Fig.~\ref{fig:UPlotC2}, displays a much larger $\xi$-dependent separation than its canonical counterpart. This amplification is a direct consequence of the Fermi statistics: changing $\xi$ not only shifts the energies but also rearranges the occupation numbers near the chemical potential. At low temperature, the internal energy is dominated by the filled states below the Fermi edge, whose total energy is strongly coupling dependent. At higher temperatures, the curves develop a common slope controlled by the Sommerfeld-like thermal broadening of the Fermi distribution~\cite{PathriaBeal2011}.
\begin{figure}[h!]
\centering
\includegraphics[width=8cm,height=5cm]{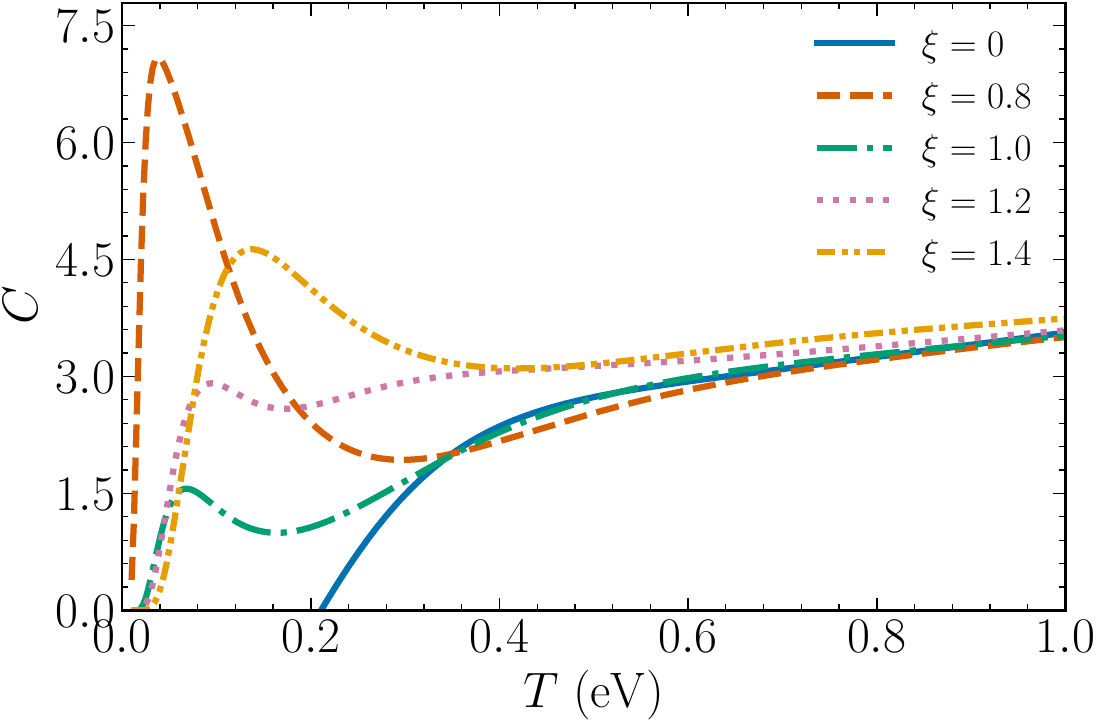}
\caption{Grand-canonical heat capacity as a function of temperature for representative couplings. The rapid low-temperature rise followed by a broad saturation region is characteristic of a mesoscopic Fermi system in which a finite cluster of levels around the chemical potential becomes thermally active within a narrow temperature window.}
\label{fig:CPlotC2}
\end{figure}

Figure~\ref{fig:CPlotC2} reveals that the grand-canonical heat capacity rises steeply at low temperature and then saturates to a broad plateau whose height depends on $\xi$. The steep rise indicates the onset of thermal activation of levels near the Fermi edge: the narrower their spacing (controlled by $\xi$), the lower the temperature at which this activation begins. The plateau value reflects the effective number of states contributing to the energy fluctuations, and its $\xi$-dependence demonstrates that the spin--orbit splitting controls the local density of states at the Fermi level.
\begin{figure}[h!]
\centering
\includegraphics[width=8cm,height=5cm]{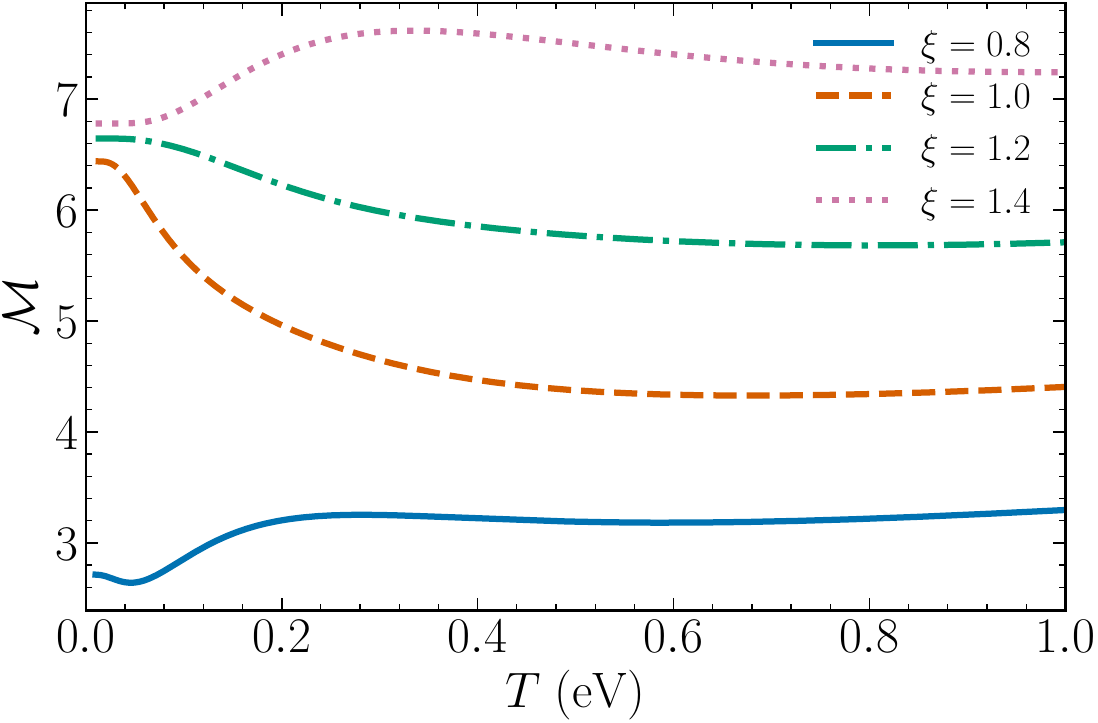}
\caption{Grand-canonical magnetization analog as a function of temperature for representative values of $\xi$. The nonmonotonic profiles arise from the competition between the direct spectral sensitivity to $\xi$ and the thermally induced redistribution of Fermi occupations.}
\label{fig:MPlotC2}
\end{figure}

Compared with the canonical case, the grand-canonical magnetization analog in Fig.~\ref{fig:MPlotC2} exhibits a substantially richer response. The nonmonotonic behavior visible for several coupling values originates from the interplay between two competing effects: the explicit dependence of the spectrum on $\xi$, which favors a monotonic response, and the temperature-driven redistribution of occupation numbers across branches that cross or approach the Fermi level~\cite{TanInkson1999}. When these branches carry opposite coupling derivatives, the two contributions partially cancel and generate local extrema in $M(T)$, a feature that is entirely absent in the canonical ensemble.
\begin{figure}[h!]
\centering
\includegraphics[width=8cm,height=5cm]{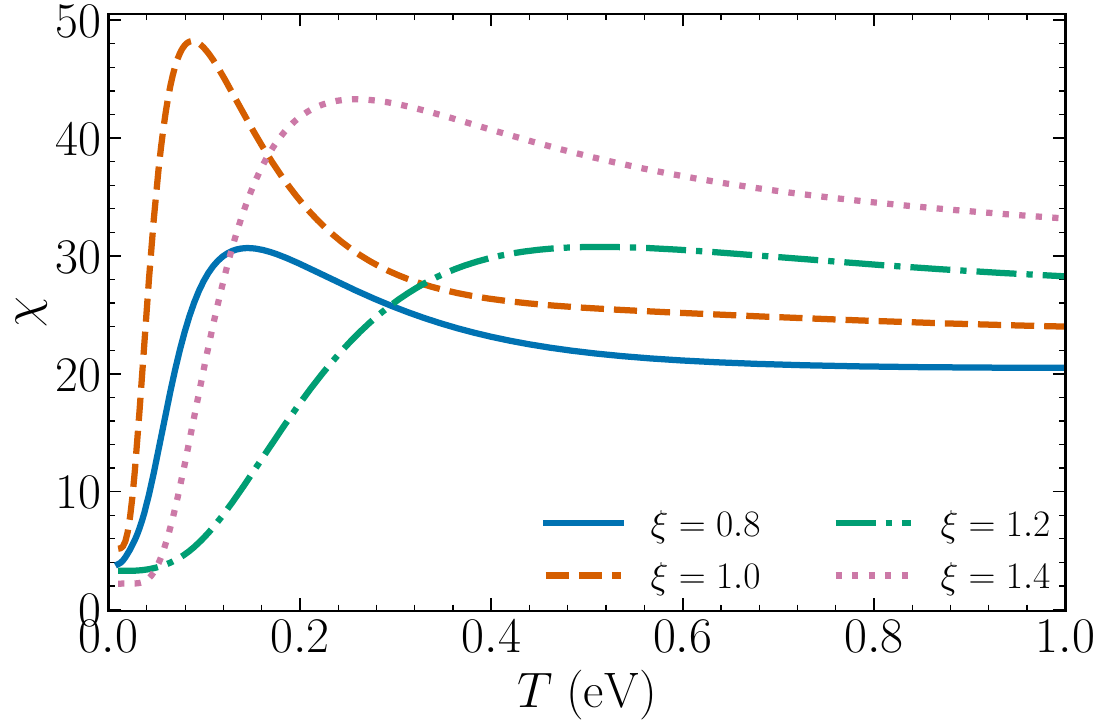}
\caption{Grand-canonical susceptibility analog as a function of temperature for representative couplings. The strong low-temperature peaks and their rapid growth with decreasing $\xi$ reflect the enhanced sensitivity of the Fermi-edge occupation to small variations of the effective spin--orbit strength.}
\label{fig:XPlotC2}
\end{figure}

The grand-canonical susceptibility in Fig.~\ref{fig:XPlotC2} confirms that derivative observables are substantially amplified by Fermi statistics. The susceptibility can reach values that are an order of magnitude larger than in the canonical case, and for certain couplings it develops a strongly peaked structure or even changes sign. This behavior is a hallmark of mesoscopic Fermi systems in which the chemical potential sits close to a coupling-dependent level crossing~\cite{CheungGefenRiedel1988,AmbegaokarEckern1990}: a small perturbation in $\xi$ then shifts a level through the Fermi edge, causing an abrupt change in occupation that is magnified in the second derivative of the grand potential.
\bigskip

\begin{figure}[h!]
\centering
\includegraphics[width=8cm,height=5cm]{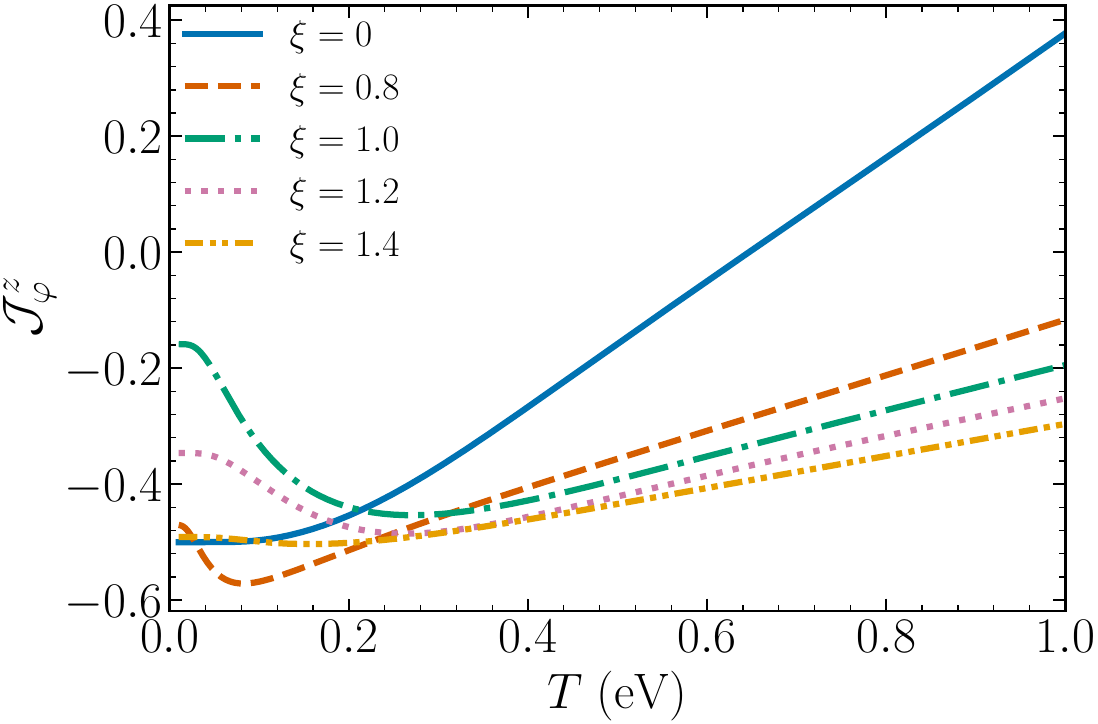}
\caption{Grand-canonical thermal expectation value of the $z$ component of the persistent spin current at fixed chemical potential. The curves demonstrate that Fermi occupation of the split branches can enhance, suppress, or even reverse the direction of the net spin current depending on temperature and effective coupling.}
\label{fig:JZgrand}
\end{figure}

\noindent\textbf{Spin current in the grand canonical ensemble.}  
The thermal expectation value of the $z$-component of the persistent spin current becomes
\begin{equation}
\langle \mathcal{J}_{\varphi}^{z} \rangle
 = \sum_{n=0}^{\infty}
   \sum_{s,\lambda=\pm}
   \frac{1}{4mr_{0}}
   \frac{ 2n\cos\theta - 1 }
        { \exp[\beta(E_{n,s}^{\lambda}-\mu)] + 1 }.
\end{equation}
Its thermal profile is shown in Fig.~\ref{fig:JZgrand}.

The grand-canonical spin current shown in Fig.~\ref{fig:JZgrand} reveals a remarkably rich temperature dependence. For $\xi=0$, the current varies nearly linearly with temperature as a growing number of angular-momentum states contribute with the same spin projection. For finite coupling, the spin texture is canted, and the contributions of different branches carry different---and in some cases opposite---spin-current signs. As the temperature increases and additional branches become occupied, partial cancellations can drive the net current through zero and even reverse its sign. This sign change is a purely many-body Fermi-surface effect: it occurs because thermally activated states above the chemical potential carry a spin current that opposes the contribution of the filled states below~\cite{Splettstoesser2003,FoersterBuettiker2008}. Such behavior has no analog in the canonical ensemble, where each ring hosts a single electron, and no competition between occupied branches can arise.

\section{Phenomenological interacting extension}
\label{Sec:InteractingApproach}

In this section, we investigate how the thermodynamic properties of a system of quantum rings are modified when weak collective effects are taken into account. Our aim is not to derive a specific microscopic interaction from the underlying relativistic theory, but rather to construct a phenomenological extension that remains analytically tractable while capturing the dominant impact of inter-ring correlations on thermodynamic observables.

In situations where a fully microscopic many-body treatment is unavailable or analytically prohibitive, it is common to describe interaction effects via effective or mean-field-like constructions at the level of thermodynamic potentials. Following this general strategy, we assume that the leading influence of weak correlations can be encoded through a deformation of the single-ring free energy, rather than through an explicit two-body Hamiltonian.

To implement this idea, we introduce an effective description in which collective effects are parameterized by a function of a dimensionless variable $\omega$ that combines temperature and the effective Rashba-like coupling. In practice, this variable will be chosen such that the interacting contribution remains anchored to the non-interacting thermodynamics, ensuring that no additional energy scale is introduced into the problem.

Within this framework, the interacting free energy will be constructed as a nonlinear functional of the non-interacting free energy. This approach allows us to systematically incorporate interaction-induced corrections via a controlled expansion while preserving the essential spectral and thermodynamic structure of the single-ring system. As will be shown below, this construction naturally leads to a resummation scheme in which collective effects manifest as nonlinear amplifications of the underlying thermodynamic response.

The interacting free energy per ring is introduced as a phenomenological nonlinear deformation of the non-interacting free energy, designed to mimic the cumulative effect of weak collective correlations in an array of quantum rings while preserving analytical tractability. Rather than attempting a microscopic cluster expansion, we assume that the dominant interaction-induced corrections can be effectively encoded through a power-series expansion of an interaction function $\phi(\omega)$, where $\omega$ denotes a collective variable depending on temperature and on the effective Rashba-like coupling.

To ensure dimensional consistency, the expansion must be written in terms of the dimensionless combination $\beta\,\phi(\omega)$. We therefore define the interacting free energy as
\begin{equation}
\mathcal{F}_{I}(T,\xi)
=
\frac{1}{\beta}
\sum_{k=1}^{\infty}
\frac{1}{k!}
\bigl[\beta\,\phi(\omega)\bigr]^k .
\label{FI_series}
\end{equation}
This construction should be interpreted as an effective resummation ansatz: the first term reproduces the leading contribution, while the higher-order terms encode progressively nonlinear collective corrections. Since the coefficients are factorially suppressed, the series is manifestly convergent for all finite values of $\beta\,\phi(\omega)$.

At the lowest order, consistency with the previously derived non-interacting problem requires that the interaction function reduce to the single-ring Helmholtz free energy. We therefore impose the self-consistency condition
\begin{equation}
\phi(\omega)=\mathcal{F}_{0}(T,\xi),
\label{selfconsistency_phi}
\end{equation}
where $\mathcal{F}_{0}(T,\xi)$ is obtained from the single-ring partition function,
\begin{equation}
\mathcal{F}_{0}(T,\xi)
=
-\frac{1}{\beta}\ln \mathcal{Z}_{1}(T,\xi).
\end{equation}
This identification anchors the interacting deformation to the same $\xi$-dependent thermodynamic structure already present in the non-interacting model, so that the phenomenological interaction does not introduce an unrelated energy scale, but rather reshapes the original free-energy landscape through a controlled nonlinear mapping.

Substituting Eq.~\eqref{selfconsistency_phi} into Eq.~\eqref{FI_series}, one obtains
\begin{equation}
\mathcal{F}_{I}(T,\xi)
=
\frac{1}{\beta}
\sum_{k=1}^{\infty}
\frac{1}{k!}
\bigl[\beta\,\mathcal{F}_{0}(T,\xi)\bigr]^k .
\label{FI_series_F0}
\end{equation}
Since the series in Eq.~\eqref{FI_series_F0} is simply the Taylor expansion of the exponential function with the zeroth-order term removed, the interacting free energy can be written in closed form as
\begin{equation}
\mathcal{F}_{I}(T,\xi)
=
\frac{1}{\beta}
\left[
\exp\!\bigl(\beta\,\mathcal{F}_{0}(T,\xi)\bigr)-1
\right].
\label{FI_closed}
\end{equation}
Equation~\eqref{FI_closed} provides a compact nonlinear resummation of the non-interacting free energy and makes explicit that the strength of the effective collective correction is controlled by the dimensionless parameter $\beta\,\mathcal{F}_{0}$. In this way, temperature plays a direct role in regulating the relevance of higher-order terms: at high temperatures, where $|\beta\,\mathcal{F}_{0}|\ll 1$, the interacting free energy approaches the non-interacting one, whereas at low temperatures, nonlinear corrections become increasingly important.

Indeed, in the regime $|\beta\,\mathcal{F}_{0}|\ll 1$, Eq.~\eqref{FI_series_F0} may be truncated at low order, yielding
\begin{equation}
\mathcal{F}_{I}(T,\xi)
\approx
\mathcal{F}_{0}(T,\xi)
+
\frac{\beta}{2}\,\mathcal{F}_{0}(T,\xi)^2
+
\frac{\beta^{2}}{6}\,\mathcal{F}_{0}(T,\xi)^3
+\cdots .
\label{FI_truncated}
\end{equation}
This truncated expansion offers a transparent physical interpretation: the first term reproduces the non-interacting contribution, while the quadratic and cubic terms represent the leading collective corrections induced by the phenomenological interaction. The truncation is therefore justified whenever the effective expansion parameter $\beta\,\mathcal{F}_{0}$ remains sufficiently small, whereas the full exponential form in Eq.~\eqref{FI_closed} provides a nonperturbative extension valid for arbitrary finite values of $\beta\,\mathcal{F}_{0}$.
The figures in this section are obtained from direct numerical evaluation of Eq.~\eqref{FI_closed}, which remains well defined for all finite temperatures and couplings; the truncated form~\eqref{FI_truncated} is used only for analytical interpretation.

The interacting ansatz introduced above should thus be viewed as a mathematically consistent and phenomenologically controlled resummation scheme. It is not intended to represent the exact cluster expansion of a microscopic many-body Hamiltonian, but rather to capture, in an analytically manageable form, how weak collective effects may amplify and reshape the thermodynamic response of the Rashba-like quantum-ring system.

The nonlinear construction introduced above should be understood within the general framework of phenomenological and mean-field approaches to interacting systems, in which the free energy is treated as an effective functional that encodes collective behavior beyond a strictly microscopic description. In particular, it is well established that, when an exact many-body treatment is not analytically tractable, the thermodynamic potential may be expanded or deformed in a manner consistent with general thermodynamic principles, symmetry constraints, and limiting behaviors, as exemplified by Landau theory of phase transitions~\cite{Landau1937,LandauLifshitzStatPhys}.

More broadly, mean-field and effective-field theories routinely construct nonlinear functionals of macroscopic quantities in order to incorporate interaction effects in an averaged or self-consistent manner~\cite{Goldenfeld1992,ChaikinLubensky}. In such approaches, the free energy is not derived from a convergent cluster expansion but instead modeled through controlled approximations that preserve thermodynamic consistency while capturing the dominant collective features of the system.

From a complementary perspective, nonlinear structures of the free energy also naturally arise in resummation schemes and cumulant-based expansions, where exponential or generating-functional forms encode higher-order correlations in a compact way~\cite{KardarStatPhys,NegeleOrland}. These constructions provide a formal basis for introducing nonlinear mappings of thermodynamic potentials when the underlying microscopic correlations are only partially accessible.

Motivated by these general principles, we adopt here a phenomenological ansatz in which the interacting free energy is constructed as a nonlinear functional of the non-interacting free energy. This choice ensures that (i) the correct non-interacting limit is recovered, $\mathcal{F}_{I}\to\mathcal{F}_{0}$ for $|\beta\mathcal{F}_{0}|\ll 1$, (ii) no additional energy scale is introduced beyond those already present in the system, and (iii) interaction effects are incorporated through a mathematically well-defined and thermodynamically consistent deformation. Within this framework, the exponential mapping can be interpreted as a controlled resummation of higher-order contributions, providing a compact and analytically tractable description of collective effects in the regime where a full microscopic treatment is not available.

Now, differentiating the identity above with respect to the thermodynamic variables yields the interacting thermodynamic functions,
\begin{align*}
S_{I} &= -\left(
            \frac{\partial \mathcal{F}_{I}}{\partial T}
         \right)_{\xi},
&
U_{I} &= \mathcal{F}_{I} + T S_{I},
\\[0.1cm]
C_{I} &= \left(
            \frac{\partial U_{I}}{\partial T}
         \right)_{\xi},
&
M_{I} &= -\left(
            \frac{\partial \mathcal{F}_{I}}{\partial \xi}
         \right)_{T}.
\end{align*}
These definitions transfer the nonlinearity of the free-energy ansatz to all response quantities, making derivative observables such as the heat capacity and susceptibility especially sensitive to the phenomenological coupling in the low-temperature regime.

These interacting quantities exhibit nontrivial features even when the interaction is parametrically small, due to the nonlinearity of the Rashba-like coupling and the non-quadratic nature of the single-particle spectrum. Their behavior is illustrated in the figures below.

\begin{figure}[h]
\centering
\includegraphics[width=8cm,height=5cm]{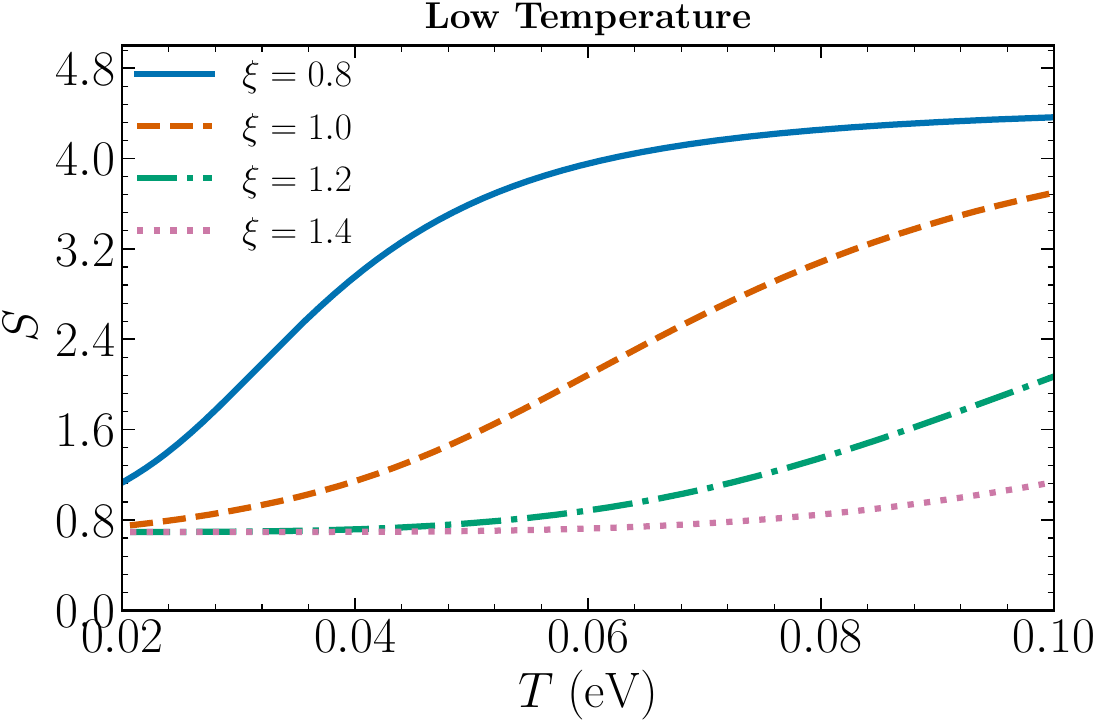}
\caption{Low-temperature interacting entropy as a function of temperature for representative values of $\xi$. The strong separation between curves demonstrates that the phenomenological interaction reshapes the thermal activation profile of the low-lying split states in a coupling-dependent manner.}
\label{fig:Inter-S}
\end{figure}

The interacting entropy in Fig.~\ref{fig:Inter-S} reveals that collective effects can dramatically alter the thermal activation landscape. Compared with the non-interacting case, where the onset of entropy growth is relatively smooth, the interacting model produces a strongly $\xi$-dependent crossover: for small couplings the entropy rises rapidly as the interaction facilitates access to nearby states, while for larger $\xi$ the onset is delayed because the effective splitting pushes the first available states to higher energies. This selectivity demonstrates that the phenomenological coupling amplifies the spectral structure, making the thermodynamic response much more sensitive to the details of the low-energy manifold.
\begin{figure}[h]
\centering
\includegraphics[width=8cm,height=5cm]{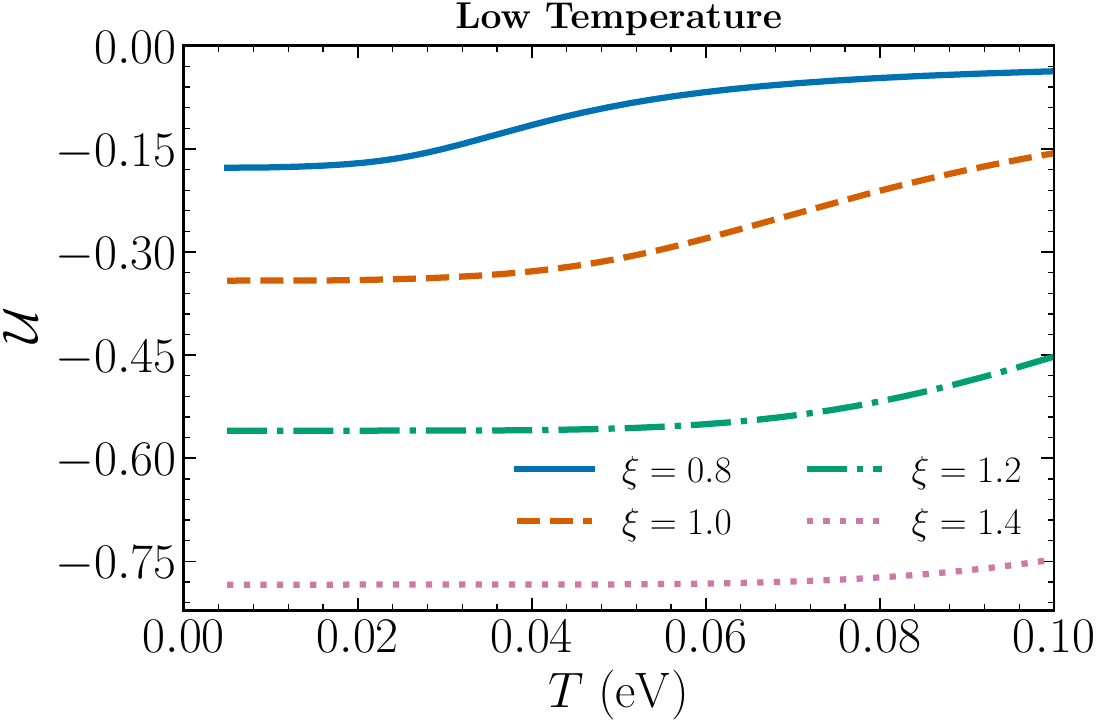}
\caption{Low-temperature interacting internal energy for representative values of $\xi$. The different slopes and offsets reflect how the phenomenological interaction modifies not only the effective ground-state energy but also the curvature of the free-energy landscape near the minimum.}
\label{fig:Inter-U}
\end{figure}

In Fig.~\ref{fig:Inter-U}, the internal energy displays a qualitative change with respect to the non-interacting case: some couplings produce a nearly flat low-temperature plateau, while others generate a steep rise already at very low $T$. This diversity indicates that the interaction reshapes the free-energy surface in a way that is sensitive to $\xi$, altering both the equilibrium energy and the effective thermal stiffness of the system. The physical implication is that an array of weakly coupled rings can exhibit a wide range of low-temperature behaviors controlled by a single spin-orbit parameter.
\begin{figure}[h]
\centering
\includegraphics[width=8cm,height=5cm]{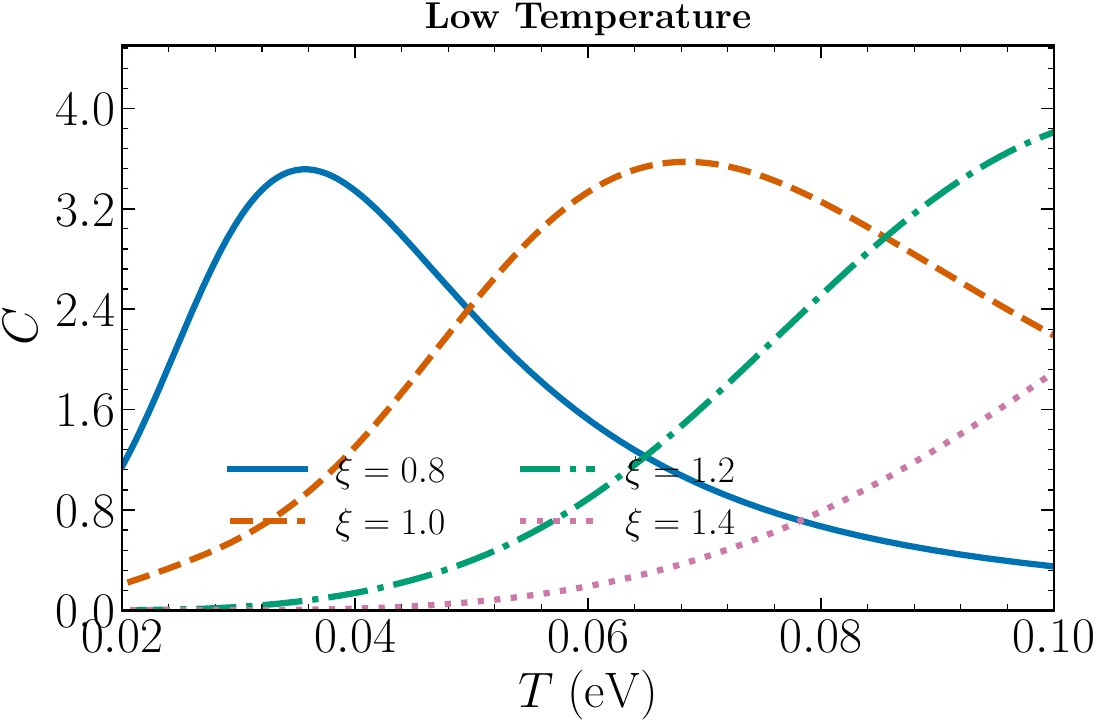}
\caption{Low-temperature interacting heat capacity for representative values of $\xi$. The large positive peaks and the strongly coupling-dependent profiles demonstrate that collective effects amplify the thermal response of the split mesoscopic spectrum far beyond the non-interacting prediction.}
\label{fig:Inter-C}
\end{figure}

The interacting heat capacity, shown in Fig.~\ref{fig:Inter-C}, is the most striking manifestation of collective amplification. The peaks are not only much larger than their non-interacting counterparts but also shift systematically with $\xi$, reflecting the coupling-dependent reorganization of the effective spectrum. Within the present phenomenological framework, the enhanced amplitude should be interpreted as evidence that the nonlinear free-energy ansatz concentrates the thermal response into a narrow temperature window, rather than as a quantitative microscopic prediction. Nevertheless, the qualitative message is clear: even weak inter-ring interactions can convert a modest Schottky anomaly into a sharp, easily detectable feature.
\begin{figure}[h]
\centering
\includegraphics[width=8cm,height=5cm]{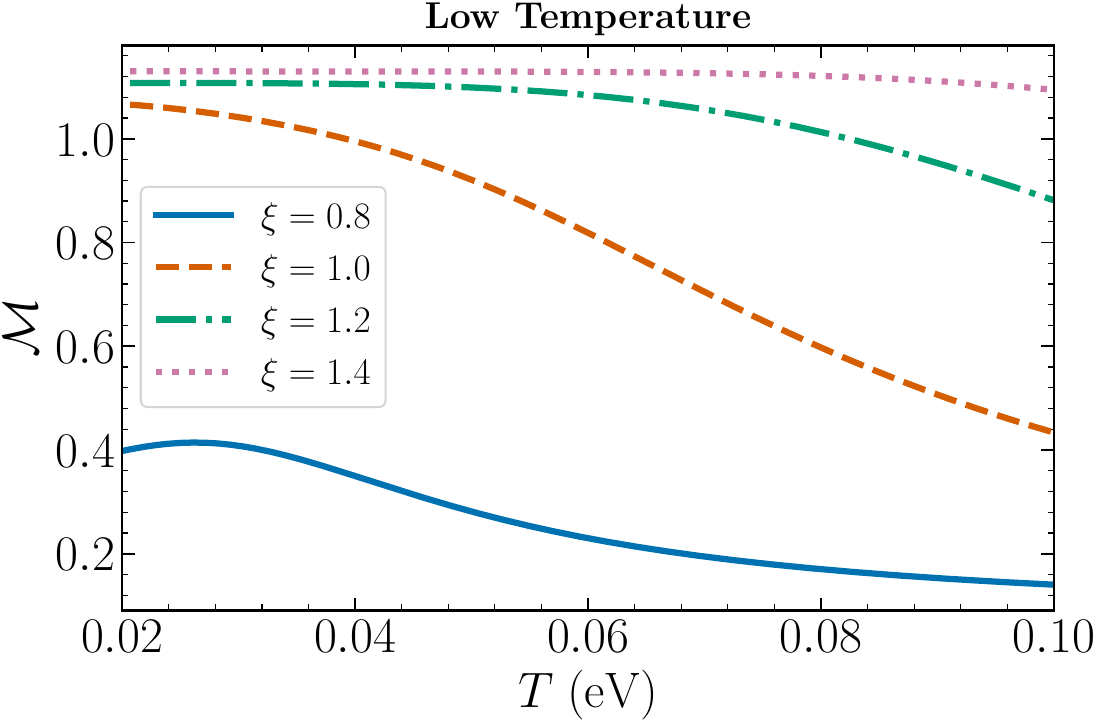}
\caption{Low-temperature interacting magnetization analog as a function of temperature for representative values of $\xi$. Depending on the coupling, the response to the effective Rashba parameter can remain nearly constant or exhibit a pronounced peak, showing that collective effects reshape the relative stability of the split branches.}
\label{fig:Inter-M}
\end{figure}

Figure~\ref{fig:Inter-M} shows that the interacting magnetization analog can develop a pronounced peak structure that is absent in the non-interacting case. For some couplings, the response remains nearly rigid, indicating that the interaction locks the system into a configuration whose free energy is insensitive to small variations of $\xi$. For others, a steep rise and subsequent decay reveal a temperature-driven transition between configurations that respond very differently to the spin--orbit parameter. This behavior is consistent with the picture in which collective effects selectively stabilize or destabilize the spin-split branches.

\begin{figure}[h]
\centering
\includegraphics[width=8cm,height=5cm]{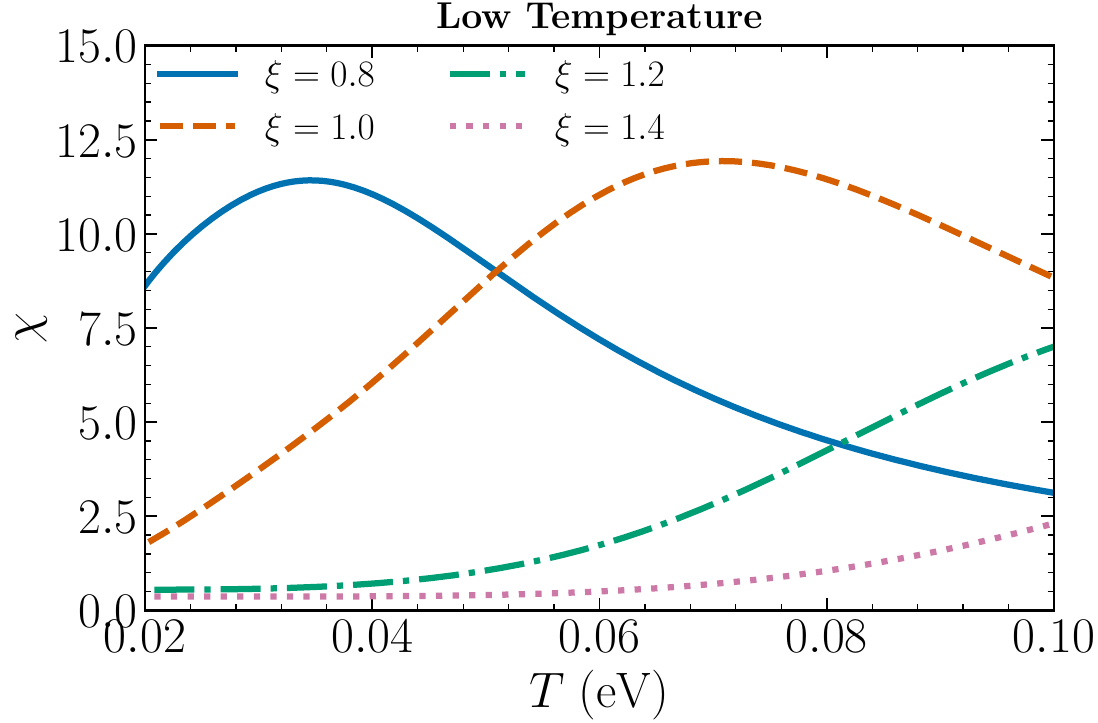}
\caption{Low-temperature interacting susceptibility analog for representative couplings. The strong enhancement compared with the non-interacting case demonstrates that collective effects can magnify the derivative response to the effective spin--orbit strength by more than an order of magnitude.}
\label{fig:Inter-X}
\end{figure}

The interacting susceptibility in Fig.~\ref{fig:Inter-X} is dramatically enhanced relative to the non-interacting result, with peak values that can exceed those of the canonical case by an order of magnitude or more. This amplification is a direct consequence of the nonlinear resummation encoded in the interacting free-energy ansatz: the exponential-resummation structure concentrates the sensitivity to $\xi$ into the temperature range where the underlying split spectrum is most structured. In practical terms, the system becomes exquisitely responsive to small changes in the effective Rashba strength once collective effects are present, suggesting that weakly coupled ring arrays could serve as sensitive probes of spin--orbit interactions.

The interacting approach developed here, therefore, provides a flexible phenomenological framework for incorporating collective or environmental effects in arrays of Rashba-like quantum rings. Even when the interaction is weak, the resulting thermodynamic response may be substantially altered, especially at low temperatures where the effective density of states is most sensitive to geometric and spin-orbit contributions. A fully microscopic interacting treatment, derived directly from the underlying relativistic theory, is left for future work.

\section{Thermomechanical response of the ring system}
\label{Sec:ThermomechanicalResponse}

In addition to the standard thermodynamic analysis developed in the previous sections, the present model also allows one to investigate effective mechanical response properties when the system's geometric size is treated as a thermodynamic variable. For a quantum ring, the natural one-dimensional analogue of the volume is the circumference
\begin{equation}
L = 2\pi r_0,
\end{equation}
which encodes the geometric scale entering the single-particle spectrum.

A central idea of this section is that the ring radius is not promoted to a dynamical variable in the Hamiltonian itself. Rather, the dependence of the spectrum on $r_0$ is explicit and parametric, which makes it possible to define quasi-static mechanical derivatives. In practice, we consider infinitesimal adiabatic deformations,
\begin{equation}
r_0 \rightarrow r_0 + \delta r_0,
\qquad
\delta r_0 \ll r_0,
\end{equation}
such that the system remains in instantaneous thermal equilibrium and no transitions between eigenstates are induced. Within this interpretation, derivatives with respect to $L$ must be understood as parametric derivatives of the energy levels and of the associated thermodynamic potentials. This procedure is directly analogous to the usual treatment of lattice constants in solid-state systems, where effective elastic and thermal coefficients are extracted from band structures computed at fixed geometry.

Throughout this section, $\xi$ is treated as an independent control parameter when differentiating with respect to~$L$. This is consistent with the physical interpretation of $\xi = mr_0\mathcal{F}_{12}$ in the quasi-static limit: maintaining $\xi$ fixed as $L$ varies amounts to assuming that the background field $\mathcal{F}_{12}$ adjusts proportionally to $1/r_0$, so that the effective spin--orbit strength remains constant throughout the deformation. This is the natural assumption when $\xi$ plays the role of an externally controlled parameter, in direct analogy with gate-voltage-tuned Rashba coupling in semiconductor heterostructures~\cite{Nitta1997,Grundler2000}.

The usefulness of this viewpoint is that it allows one to define pressure-like observables and response coefficients that emerge directly from the quantum spectrum. Because the spectrum depends on $L$ through the geometric prefactor
\begin{equation}
\Omega = \frac{1}{2mr_0^2} = \frac{2\pi^2}{mL^2},
\end{equation}
all thermomechanical quantities are ultimately controlled by the same spectral deformation that governs the thermal observables discussed in Sec.~\ref{Sec:Thermo-approach}. In this sense, the present section provides a direct bridge between the microscopic dependence of the energy levels on the ring size and the system's macroscopic effective response.

For the numerical illustrations shown below, we evaluate the exact discrete spectrum directly, keeping the full summation over the mesoscopic levels. The plots were generated in natural units with $m=1$ and $L=2\pi$ (equivalently $r_0=1$). The coupling values are those displayed in each figure legend: broad canonical plots use representative values such as $\xi=0$, $0.8$, $1.0$, $1.2$, and $1.4$, whereas the thermomechanical figures use finer coupling grids whenever this improves the visibility of crossings, divergences, or peak shifts. In the grand-canonical sector, the chemical potential is fixed at $\mu=1$. These choices do not affect the formal derivations, but they provide a convenient and transparent reference scale for visualizing the thermomechanical response.

\subsection{Canonical ensemble}
\label{Sec:ThermomechCanonical}

In the canonical ensemble, the relevant thermodynamic potential is the Helmholtz free energy per ring,
\begin{equation}
\mathcal{F}(T,L,\xi) = -\frac{1}{\beta}\ln \mathcal{Z}_1,
\end{equation}
where $\mathcal{Z}_1$ is constructed from the single-particle spectrum
\begin{equation}
E_{n,s}^{\lambda}
=
\Omega
\left[
\left(
n + \frac{1}{2}
- \frac{\lambda s}{2}\sqrt{1+4\xi^2}
\right)^2
- \xi^2
\right].
\end{equation}
At fixed $\xi$, the entire $L$-dependence enters through $\Omega\propto L^{-2}$, which means that the geometric variation of the spectrum is homogeneous. This simple scaling property has an important consequence: the effective pressure-like variable can be written in closed form in terms of the internal energy.

We define the effective one-dimensional pressure by
\begin{equation}
\mathcal{P}
=
-\left(\frac{\partial \mathcal{F}}{\partial L}\right)_{T,\xi}.
\end{equation}
Because the spectrum scales as $L^{-2}$, one finds
\begin{equation}
\mathcal{P} = \frac{2U}{L},
\label{eq:P-canonical-final-rev}
\end{equation}
where $U$ is the canonical internal energy. This relation is particularly transparent: the effective pressure is not an independent quantity but rather a direct measure of the internal energy density associated with the geometric confinement.

The thermal pressure coefficient is then
\begin{equation}
\beta_L
=
\left(\frac{\partial \mathcal{P}}{\partial T}\right)_{L,\xi}
=
\frac{2C}{L},
\label{eq:beta-canonical-final-rev}
\end{equation}
with $C$ the heat capacity. Therefore, the thermal sensitivity of the pressure-like response is directly governed by the same energy fluctuations that determine the canonical heat capacity.

The isothermal compressibility is defined as
\begin{equation}
\kappa_T
=
-\frac{1}{L}
\left(\frac{\partial L}{\partial \mathcal{P}}\right)_{T,\xi},
\end{equation}
which may be expressed as
\begin{equation}
\kappa_T
=
-\frac{1}{L}
\left[
\frac{2}{L}\left(\frac{\partial U}{\partial L}\right)_{T,\xi}
-\frac{2U}{L^2}
\right]^{-1}.
\label{eq:kappa-canonical-final-rev}
\end{equation}
Finally, the analog of the isobaric expansion coefficient is
\begin{equation}
\alpha_{\mathcal{P}}
=
\frac{1}{L}
\left(\frac{\partial L}{\partial T}\right)_{\mathcal{P},\xi}
=
\kappa_T\,\beta_L.
\label{eq:alpha-canonical-final-rev}
\end{equation}
This identity shows that the expansion response is controlled by the interplay between thermal fluctuations and mechanical softness.
\begin{figure}[t]
\centering
\includegraphics[width=0.95\columnwidth]{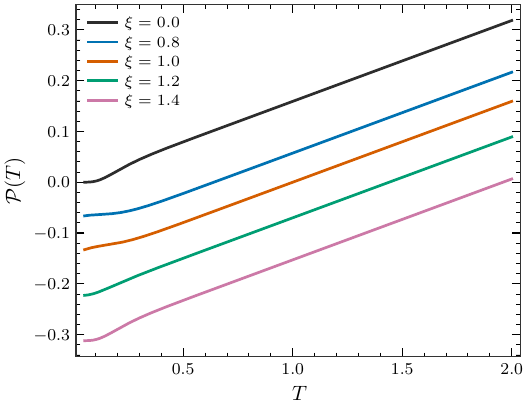}
\caption{Canonical effective pressure $\mathcal{P}(T)$ as a function of temperature for representative values of the effective coupling $\xi$. The curves were obtained from the exact discrete spectrum using $\mathcal{P}=2U/L$, with natural units $m=1$ and $L=2\pi$. Finite $\xi$ shifts the low-temperature baseline downward because the Rashba-like deformation lowers the internal energy of the lowest occupied branches. At higher temperatures, all curves increase smoothly as additional angular-momentum states become thermally accessible.}
\label{fig:ThermoMechP}
\end{figure}

Figure~\ref{fig:ThermoMechP} provides the first direct visualization of the canonical thermomechanical response. Since $\mathcal{P}$ is proportional to the internal energy, its temperature dependence inherits the same qualitative features already discussed for $U(T)$, but now interpreted as an effective mechanical observable. At low temperature, the pressure is determined almost entirely by the ground-state sector. Consequently, increasing $\xi$ displaces the curves downward because the spin--orbit-induced spectral reconstruction lowers the energy of the lowest branches. As the temperature rises, excited angular-momentum states begin to contribute, and the convergence trend among the curves at intermediate and high temperature indicates that the details of the low-lying branch rearrangement become less dominant once many states enter the canonical average.
\begin{figure}[t]
\centering
\includegraphics[width=0.95\columnwidth]{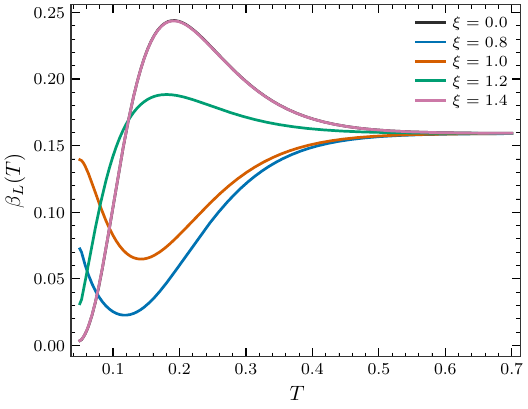}
\caption{Canonical thermal pressure coefficient $\beta_L(T)=\left(\partial\mathcal{P}/\partial T\right)_{L,\xi}=2C/L$ for representative values of $\xi$. Because $\beta_L$ is directly proportional to the heat capacity, its profile delineates the temperature range over which the redistribution of Boltzmann weights among the split mesoscopic levels is most pronounced.}
\label{fig:ThermoMechBeta}
\end{figure}

The physical content of Fig.~\ref{fig:ThermoMechBeta} is especially important because $\beta_L$ is the direct thermomechanical analog of the heat capacity. The identity $\beta_L=2C/L$ means that every thermal fluctuation contributing to the energy variance also contributes to the temperature derivative of the pressure. The nontrivial peaks and shoulders visible in the curves therefore identify the temperature scales at which the thermal population of the split branches changes most rapidly. For $\xi=0$, the coefficient exhibits the standard broad activation structure expected from a mesoscopic ladder with a single dominant low-energy scale. Once $\xi$ is turned on, the response develops richer low-temperature features because different sets of levels become thermally active at different temperatures.
\begin{figure}[t]
\centering
\includegraphics[width=0.95\columnwidth]{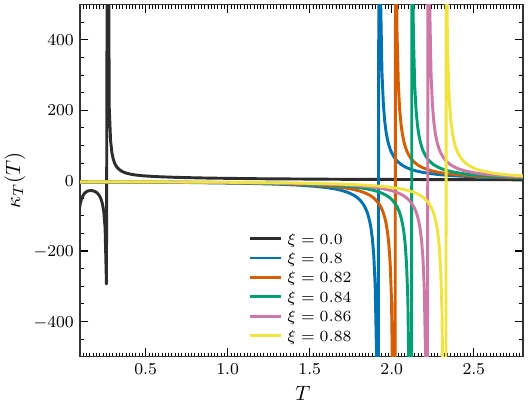}
\caption{Canonical isothermal compressibility $\kappa_T(T)$ as a function of temperature for $\xi=0.0$, $0.80$, $0.82$, $0.84$, $0.86$, and $0.88$. The quantity measures the inverse sensitivity of the effective pressure to quasi-static changes in the ring circumference at fixed temperature. The fine spacing in $\xi$ reveals that the compressibility divergences are highly sensitive to the coupling strength and occur within a narrow band of parameters, underscoring the mesoscopic origin of the mechanical instabilities.}
\label{fig:ThermoMechKappa}
\end{figure}

The isothermal compressibility shown in Fig.~\ref{fig:ThermoMechKappa} adds a genuinely mechanical perspective to the analysis. While $\mathcal{P}$ and $\beta_L$ are directly tied to $U$ and $C$, the compressibility depends on the derivative of the pressure with respect to the system size and therefore probes how rigidly the spectrum reacts to an infinitesimal deformation of the ring circumference. Different values of $\xi$ reorganize the branch structure near the ground state, which changes not only the internal energy but also the sensitivity of that energy to geometric deformation. The fine coupling spacing used here ($\Delta\xi=0.02$) highlights that the positions of the compressibility divergences are highly sensitive to $\xi$, reflecting the sharp dependence of level-crossing temperatures on the spin--orbit strength.

\begin{figure}[t]
\centering
\includegraphics[width=0.95\columnwidth]{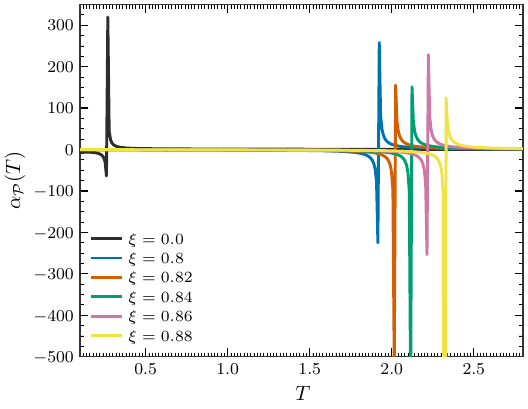}
\caption{Canonical isobaric expansion coefficient analogue $\alpha_{\mathcal{P}}(T)=\kappa_T\beta_L$ for $\xi=0.0$, $0.80$, $0.82$, $0.84$, $0.86$, and $0.88$. This coefficient condenses the full thermomechanical content of the canonical ring problem into a single observable. Its structure reveals the temperature window in which thermal activation and geometric softness cooperate most strongly, and the fine-coupling spacing illustrates the high sensitivity of the response to small variations in the spin--orbit parameter.}
\label{fig:ThermoMechAlpha}
\end{figure}

The expansion coefficient analog displayed in Fig.~\ref{fig:ThermoMechAlpha} is the most synthetic canonical thermomechanical quantity. Because $\alpha_{\mathcal{P}}=\kappa_T\beta_L$, it inherits features from both the thermal pressure coefficient and the compressibility. It becomes large precisely in the regime where two conditions are simultaneously met: the pressure must change appreciably with temperature, and the system must remain sufficiently responsive to an infinitesimal quasi-static deformation of the circumference.

\begin{figure}[t]
\centering
\includegraphics[width=0.95\columnwidth]{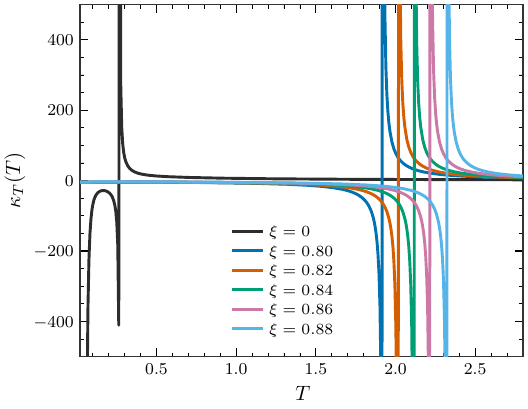}
\caption{Canonical isothermal compressibility $\kappa_T(T)$ over the displayed temperature interval for $\xi=0$, $0.80$, $0.82$, $0.84$, $0.86$, and $0.88$. The fine-coupling spacing reveals the coupling-selective nature of the divergences: each value of $\xi$ yields a compressibility peak at a distinct temperature, thereby directly mapping the positions of the level-crossing instabilities induced by the spin-orbit deformation.}
\label{fig:CanKappaZoom}
\end{figure}

\begin{figure}[t]
\centering
\includegraphics[width=0.95\columnwidth]{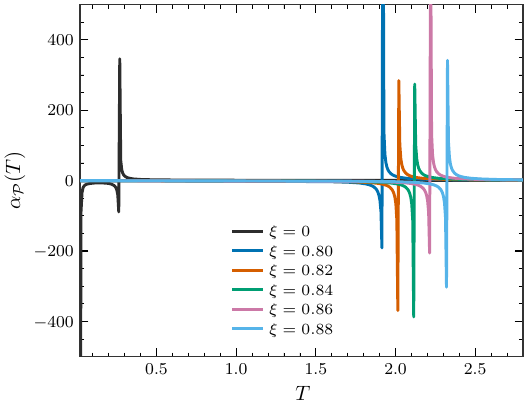}
\caption{Canonical isobaric expansion coefficient $\alpha_{\mathcal{P}}(T)$ over the displayed temperature interval for $\xi=0$, $0.80$, $0.82$, $0.84$, $0.86$, and $0.88$. The figure reveals a rich crossing pattern across different $\xi$ values, indicating that the coupling-dependent interplay between thermal fluctuations and mechanical compliance yields a nontrivial reordering of the expansion response as a function of temperature.}
\label{fig:CanAlphaZoom}
\end{figure}

To resolve the fine coupling dependence of the canonical response, we present in Figs.~\ref{fig:CanKappaZoom} and~\ref{fig:CanAlphaZoom} additional views of $\kappa_T(T)$ and $\alpha_{\mathcal{P}}(T)$ over the temperature interval displayed in the corresponding plots.

The detailed view of the compressibility in Fig.~\ref{fig:CanKappaZoom} resolves the coupling-selective divergences clearly: each of the finely spaced $\xi$ values produces a peak at a distinct temperature, demonstrating that the level-crossing temperatures shift continuously with $\xi$ and that the compressibility constitutes a high-resolution probe of the spin-orbit deformation. The expansion coefficient in Fig.~\ref{fig:CanAlphaZoom} displays a rich pattern of crossings, indicating that the product $\kappa_T\beta_L$ is governed by a delicate balance between thermal activation and mechanical response whose relative weight is highly sensitive to both $\xi$ and $T$ in the mesoscopic low-temperature regime.

Taken together, Figs.~\ref{fig:ThermoMechP}--\ref{fig:CanAlphaZoom} show that the canonical ring exhibits a complete and internally consistent thermomechanical response. The effective pressure is controlled by the internal energy density; the thermal pressure coefficient directly tracks the heat capacity; the compressibility measures the geometric softness of the spectrum; and the expansion coefficient combines both thermal and mechanical sensitivities into a single response function. The entire structure emerges from the explicit dependence of the energy levels on the circumference and requires no additional phenomenological input beyond the exact quantum spectrum.

\subsection{Grand canonical ensemble}
\label{Sec:ThermomechGrandCanonical}

In the grand canonical ensemble, the relevant thermodynamic potential is
\begin{equation}
\Phi(T,\mu,L,\xi)
=
-\frac{1}{\beta}
\sum_{n,s,\lambda}
\ln\left[1+e^{-\beta(E_{n,s}^{\lambda}-\mu)}\right].
\end{equation}
The corresponding effective pressure is defined by
\begin{equation}
\mathcal{P}
=
-\left(\frac{\partial \Phi}{\partial L}\right)_{T,\mu,\xi},
\end{equation}
which yields
\begin{equation}
\mathcal{P}
=
\frac{2}{L}
\sum_{n,s,\lambda}
E_{n,s}^{\lambda}\,
f(E_{n,s}^{\lambda}),
\end{equation}
where
\begin{equation}
f(E)=\frac{1}{e^{\beta(E-\mu)}+1}
\end{equation}
is the Fermi--Dirac occupation function. Introducing the grand-canonical internal energy,
\begin{equation}
U =
\sum_{n,s,\lambda}
E_{n,s}^{\lambda}\, f(E_{n,s}^{\lambda}),
\end{equation}
one recovers the same formal relation as in the canonical ensemble,
\begin{equation}
\mathcal{P} = \frac{2U}{L}.
\end{equation}
The associated response coefficients are therefore
\begin{equation}
\beta_L = \frac{2}{L}
\left(\frac{\partial U}{\partial T}\right)_{L,\mu,\xi},
\end{equation}
\begin{equation}
\kappa_T
=
-\frac{1}{L}
\left[
\frac{2}{L}\left(\frac{\partial U}{\partial L}\right)_{T,\mu,\xi}
-\frac{2U}{L^2}
\right]^{-1},
\end{equation}
and
\begin{equation}
\alpha_{\mathcal{P}} = \kappa_T\,\beta_L.
\end{equation}

Although the formal thermomechanical structure is identical to the canonical one, the physical content is substantially richer. In the grand canonical description, the chemical potential determines which split branches lie close to the Fermi edge, and this makes all response coefficients highly sensitive to occupation rearrangements. A small change in $\xi$ may shift a branch across the Fermi level, which modifies not only the internal energy itself but also its derivatives with respect to temperature and circumference. As a result, one expects stronger oscillatory structures, enhanced peaks, and possibly sign changes in derivative observables, especially in the low-temperature regime.

\begin{figure}[t]
\centering
\includegraphics[width=0.95\columnwidth]{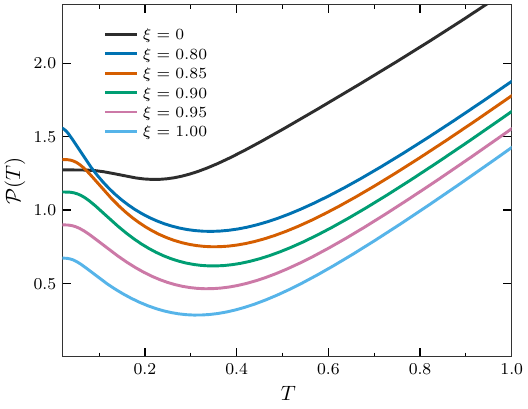}
\caption{Grand-canonical effective pressure $\mathcal{P}(T)$ as a function of temperature for $\mu=1$ and $\xi=0$, $0.80$, $0.85$, $0.90$, $0.95$, and $1.00$. Unlike the canonical case, the Fermi--Dirac occupation generates a pronounced vertical separation between the curves even at low temperatures, reflecting the coupling-dependent redistribution of filled states below the chemical potential. The nonmonotonic behavior visible for finite couplings arises from the competition between the thermal broadening of the Fermi edge and the $\xi$-dependent rearrangement of the occupied branches.}
\label{fig:GC-P}
\end{figure}

Figure~\ref{fig:GC-P} presents the grand-canonical effective pressure. Compared with the canonical result (Fig.~\ref{fig:ThermoMechP}), the most striking difference is the much larger vertical separation between curves at low temperature. This amplification is a direct consequence of the Fermi statistics: for a given chemical potential, different values of $\xi$ place different numbers of filled states below the Fermi edge, producing widely different total energies and therefore widely different pressures. The nonmonotonic behavior visible for several coupling values, where $\mathcal{P}(T)$ initially decreases before increasing, reflects the fact that, at very low temperature, the thermal broadening of the Fermi edge initially redistributes occupation from high-energy filled states to lower-lying empty ones, temporarily reducing the total energy before the usual thermal activation eventually prevails.
\begin{figure}[t]
\centering
\includegraphics[width=0.95\columnwidth]{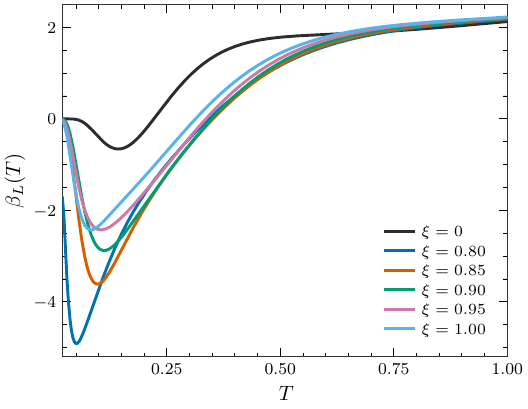}
\caption{Grand-canonical thermal pressure coefficient $\beta_L(T)=2(\partial U/\partial T)_{L,\mu,\xi}/L$ for $\mu=1$ and $\xi=0$, $0.80$, $0.85$, $0.90$, $0.95$, and $1.00$. The sharp negative excursions at low temperature identify regimes where the Fermi edge crosses a cluster of split levels, producing an anomalous thermal softening that has no counterpart in the canonical ensemble. At higher temperatures, all curves converge toward a common positive regime as the Sommerfeld-like broadening smooths out the mesoscopic level structure.}
\label{fig:GC-beta}
\end{figure}

The grand-canonical thermal pressure coefficient, shown in Fig.~\ref{fig:GC-beta}, provides a differential view of the pressure response. Unlike the canonical $\beta_L$, which remains positive throughout the displayed range, the grand-canonical coefficient develops pronounced negative excursions at low temperature for several coupling values. These negative regions correspond to temperature intervals in which heating the system temporarily \emph{decreases} the effective pressure, a genuinely many-body effect that arises when the Fermi edge lies close to a coupling-dependent cluster of split levels. In such a configuration, a small increase in temperature redistributes occupation in a way that lowers the total energy, and hence the pressure, before higher levels become thermally accessible. The depth and position of the negative excursion shift systematically across the displayed sequence $\xi=0.80$--$1.00$, showing that the thermal-pressure response is controlled by how close the Fermi edge lies to the corresponding spin-orbit-shifted level cluster.
\begin{figure}[t]
\centering
\includegraphics[width=0.95\columnwidth]{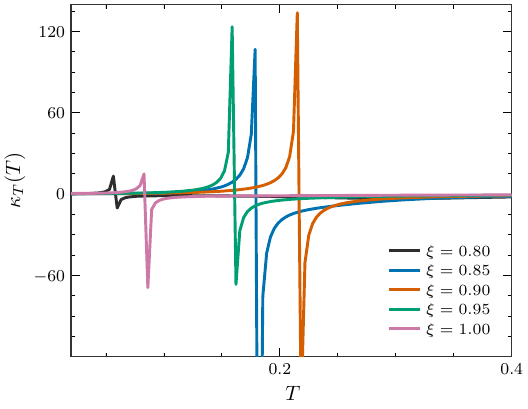}
\caption{Grand-canonical isothermal compressibility $\kappa_T(T)$ for $\mu=1$ and $\xi=0$, $0.80$, $0.85$, $0.90$, $0.95$, and $1.00$. The divergent structures correspond to temperatures at which the denominator of $\kappa_T$ passes through zero --- a signature of a coupling-dependent mechanical instability where the effective pressure becomes insensitive to infinitesimal changes in the circumference. The $\xi=0$ curve displays additional divergences outside the low-temperature cluster, reflecting the richer level-crossing structure of the uncoupled spectrum at this chemical potential.}
\label{fig:GC-kappa}
\end{figure}

The grand-canonical isothermal compressibility, displayed in Fig.~\ref{fig:GC-kappa}, reveals the most dramatic departure from canonical behavior. The divergent structures visible for each coupling correspond to temperatures at which the denominator of $\kappa_T$, proportional to $\partial\mathcal{P}/\partial L$ at fixed $T$, passes through zero, signaling a mechanical instability at which the effective pressure becomes momentarily insensitive to infinitesimal geometric deformations. These divergences are genuine Fermi-surface effects: they arise when the thermal redistribution of occupied levels exactly compensates the geometric scaling of the spectrum, producing a flat pressure-versus-length landscape at a specific temperature. Their positions are strongly $\xi$-dependent, confirming that the spin--orbit coupling controls the effective mechanical response through its reorganization of the filled branches near the chemical potential. The $\xi=0$ curve displays particularly rich structure, with multiple divergences at intermediate and high temperatures that reflect the dense level-crossing pattern of the uncoupled spectrum.

\begin{figure}[t]
\centering
\includegraphics[width=0.95\columnwidth]{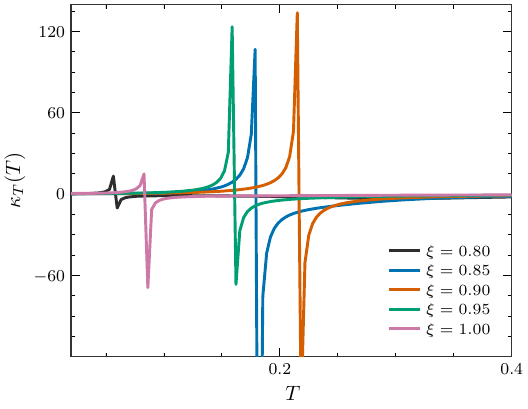}
\caption{Grand-canonical isothermal compressibility $\kappa_T(T)$ for $\mu=1$, restricted to finite couplings $\xi=0.80$, $0.85$, $0.90$, $0.95$, and $1.00$ (the $\xi=0$ curve is omitted to resolve the finer structure). The sharp peaks mark divergences associated with coupling-dependent mechanical instabilities. Their positions shift strongly with $\xi$, showing that small changes in the spin--orbit strength move the relevant Fermi-edge crossing across the displayed temperature interval.}
\label{fig:GC-kappa-noxi0}
\end{figure}

Because the $\xi=0$ curve obscures part of the finite-coupling structure in Fig.~\ref{fig:GC-kappa}, we present in Fig.~\ref{fig:GC-kappa-noxi0} the same quantity restricted to the finite-coupling values $\xi=0.80$, $0.85$, $0.90$, $0.95$, and $1.00$, in order to resolve the finer features of the compressibility.

With the $\xi=0$ curve removed, Fig.~\ref{fig:GC-kappa-noxi0} reveals a clear physical picture. All finite-coupling curves in the displayed set develop sharp structures, but the location and sign of the divergence vary strongly from $\xi=0.80$ to $1.00$. This selectivity underscores the diagnostic power of $\kappa_T$: it not only detects the presence of level crossings but also resolves their coupling dependence with high sensitivity.
\begin{figure}[t]
\centering
\includegraphics[width=0.95\columnwidth]{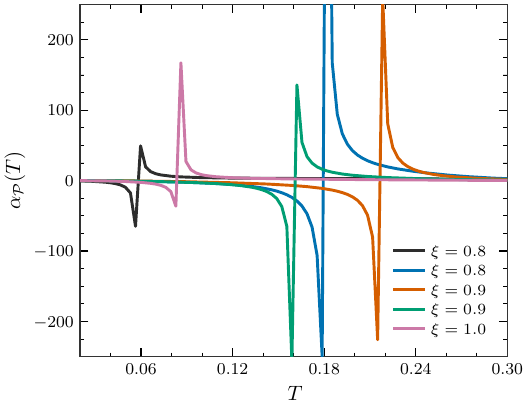}
\caption{Grand-canonical isobaric expansion coefficient $\alpha_{\mathcal{P}}(T)=\kappa_T\beta_L$ for $\mu=1$ and $\xi=0$, $0.80$, $0.85$, $0.90$, $0.95$, and $1.00$. The product inherits the divergent structures from $\kappa_T$, and the sign changes from $\beta_L$, producing a richly structured response that encapsulates the full interplay between Fermi statistics, geometric confinement, and spin--orbit coupling. The $\xi=0$ curve again displays the most extreme excursions, reflecting the mechanical instabilities of the uncoupled spectrum at this chemical potential.}
\label{fig:GC-alpha}
\end{figure}

\begin{figure}[t]
\centering
\includegraphics[width=0.95\columnwidth]{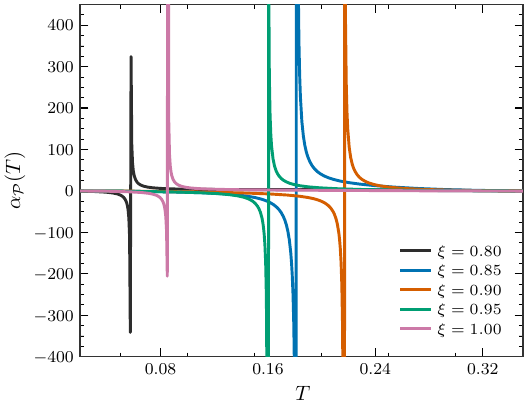}
\caption{Grand-canonical isobaric expansion coefficient $\alpha_{\mathcal{P}}(T)$ for $\mu=1$, restricted to finite couplings $\xi=0.80$, $0.85$, $0.90$, $0.95$, and $1.00$. The sharp structures originate from the same mechanical instabilities visible in $\kappa_T$, while their coupling-dependent positions and signs show how sensitively the expansion response tracks the Fermi-edge rearrangement.}
\label{fig:GC-alpha-noxi0}
\end{figure}

The grand-canonical expansion coefficient, shown in Fig.~\ref{fig:GC-alpha}, combines the features of both $\kappa_T$ and $\beta_L$ into a single response function. The divergences arising from compressibility and the sign changes from the thermal pressure coefficient produce a richly structured observable that is arguably the most sensitive probe of the thermomechanical regime. As in the compressibility case, the $\xi=0$ curve dominates the vertical scale, and we therefore present the finite-coupling expansion coefficient separately in Fig.~\ref{fig:GC-alpha-noxi0}.

Figure~\ref{fig:GC-alpha-noxi0} confirms the picture already emerging from the compressibility analysis. The expansion coefficient develops sharp peaks and sign changes whose positions move across the displayed temperature range as $\xi$ is varied from $0.80$ to $1.00$. The crossover between these structures is controlled by the proximity of the Fermi edge to the coupling-dependent level crossings: when the chemical potential sits close to a near-degeneracy in the split spectrum, the thermomechanical response is dramatically enhanced; when the Fermi edge lies farther from such a crossing, the response remains comparatively mild.

\begin{figure}[t]
\centering
\includegraphics[width=0.95\columnwidth]{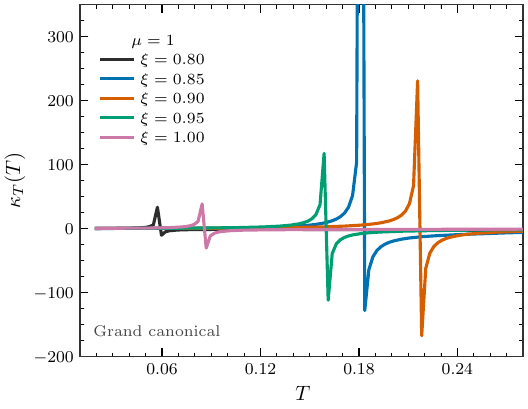}
\caption{Grand-canonical isothermal compressibility $\kappa_T(T)$ at low temperature ($T\leq 0.3$) for $\mu=1$ and finite couplings $\xi=0.80$, $0.85$, $0.90$, $0.95$, and $1.00$. The zoomed view resolves the coupling-dependent divergences and sign changes, directly mapping how the underlying level crossings in the split spectrum move as the effective spin--orbit strength is varied.}
\label{fig:GC-kappa-zoom}
\end{figure}

\begin{figure}[t]
\centering
\includegraphics[width=0.95\columnwidth]{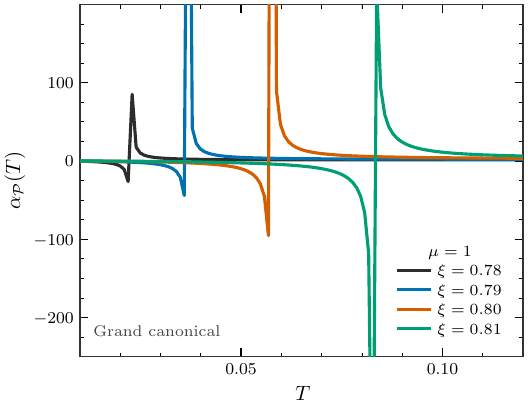}
\caption{Grand-canonical isobaric expansion coefficient $\alpha_{\mathcal{P}}(T)$ in the ultra-low-temperature window shown for $\mu=1$ and finite couplings $\xi=0.78$, $0.79$, $0.80$, and $0.81$. The zoomed view reveals that the expansion coefficient inherits the sharp divergent structures from $\kappa_T$ while acquiring additional sign changes from $\beta_L$. The result is a highly structured, coupling-selective response that serves as the most differential probe of the thermomechanical regime.}
\label{fig:GC-alpha-zoom}
\end{figure}

To further resolve the low-temperature structure of the grand-canonical response, Figs.~\ref{fig:GC-kappa-zoom} and~\ref{fig:GC-alpha-zoom} present zoomed views of $\kappa_T$ and $\alpha_{\mathcal{P}}$ using the finite-coupling values shown in their legends.


The low-temperature zooms in Figs.~\ref{fig:GC-kappa-zoom} and~\ref{fig:GC-alpha-zoom} provide the clearest picture of the grand-canonical thermomechanical instabilities. In Fig.~\ref{fig:GC-kappa-zoom}, the divergences shift across the low-temperature interval as $\xi$ is varied from $0.80$ to $1.00$. In Fig.~\ref{fig:GC-alpha-zoom}, an even finer coupling window, $\xi=0.78$--$0.81$, is used to resolve the rapid displacement of the expansion-coefficient peaks. The expansion coefficient inherits and amplifies these features through its multiplicative dependence on $\beta_L$, making it the observable with the richest low-temperature structure.

\begin{figure}[t]
\centering
\includegraphics[width=0.95\columnwidth]{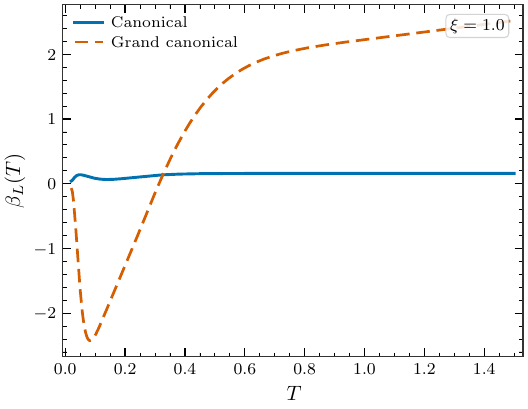}
\caption{Comparison of the thermal pressure coefficient $\beta_L(T)$ between the canonical and grand-canonical ensembles at $\xi=1.0$ ($\mu=1$ for the grand-canonical case). The canonical coefficient (solid line) remains positive and nearly flat throughout the displayed range, reflecting the smooth thermal activation of Boltzmann-weighted levels in a single-particle description. The grand-canonical coefficient (dashed line) displays a pronounced negative excursion at low temperature followed by a strong positive peak, demonstrating that Fermi statistics amplify the thermomechanical response by an order of magnitude relative to the canonical case.}
\label{fig:Compare-beta}
\end{figure}

\begin{figure}[t]
\centering
\includegraphics[width=0.95\columnwidth]{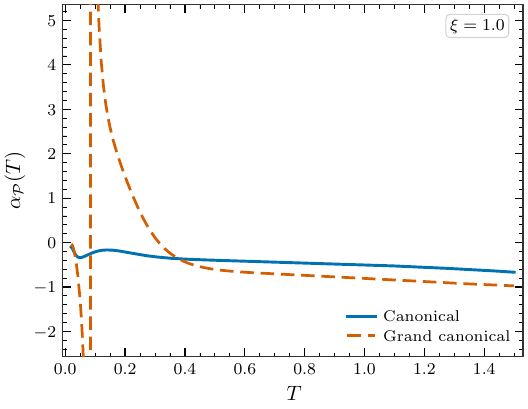}
\caption{Comparison of the isobaric expansion coefficient $\alpha_{\mathcal{P}}(T)$ between the canonical and grand-canonical ensembles at $\xi=1.0$ ($\mu=1$ for the grand-canonical case). The canonical expansion coefficient (solid line) varies slowly and remains of order unity, while the grand-canonical result (dashed line) develops a sharp divergent structure at low temperature arising from the mechanical instability identified in $\kappa_T$. This comparison provides the most vivid illustration of how Fermi-surface effects amplify the thermomechanical response of the quantum ring system.}
\label{fig:Compare-alpha}
\end{figure}

To make the comparison between ensembles explicit, Figs.~\ref{fig:Compare-beta} and~\ref{fig:Compare-alpha} overlay the canonical and grand-canonical results for the thermal pressure coefficient and the expansion coefficient at $\xi=1.0$.

The ensemble comparison in Figs.~\ref{fig:Compare-beta} and~\ref{fig:Compare-alpha} provides the most direct evidence that the grand-canonical thermomechanical response is qualitatively different from its canonical counterpart. The canonical thermal pressure coefficient varies slowly and remains of order $10^{-1}$ throughout the displayed range, while the grand-canonical coefficient develops a negative excursion reaching $\beta_L\approx -2.5$ followed by a positive peak exceeding $\beta_L\approx 2.3$. The expansion coefficient comparison is even more dramatic: the canonical $\alpha_{\mathcal{P}}$ remains bounded and nearly featureless, while the grand-canonical result diverges at the temperature corresponding to the mechanical instability. This contrast unambiguously demonstrates that Fermi statistics amplify the thermomechanical response, converting the modest signatures of the canonical problem into sharp, experimentally distinctive features.

\subsection{Phenomenological interacting extension}
\label{Sec:ThermomechInteracting}

We now extend the thermomechanical construction to the interacting model introduced in Sec.~\ref{Sec:InteractingApproach}. In this case, the thermodynamics is governed by the interacting free energy $\mathcal{F}_I(T,L,\xi)$, which encodes collective effects via the previously adopted phenomenological ansatz. The corresponding pressure-like quantity is defined by
\begin{equation}
\mathcal{P}_I
=
-\left(\frac{\partial \mathcal{F}_I}{\partial L}\right)_{T,\xi},
\end{equation}
and the associated response coefficients are
\begin{equation}
\kappa_T^{(I)}
=
-\frac{1}{L}
\left[
\left(\frac{\partial \mathcal{P}_I}{\partial L}\right)_{T,\xi}
\right]^{-1},
\end{equation}
\begin{equation}
\beta_L^{(I)}
=
\left(\frac{\partial \mathcal{P}_I}{\partial T}\right)_{L,\xi},
\end{equation}
and
\begin{equation}
\alpha_{\mathcal{P}}^{(I)}
=
\kappa_T^{(I)}\,\beta_L^{(I)}.
\end{equation}

In contrast to the noninteracting problem, the interacting coefficients are no longer determined solely by the homogeneous $L^{-2}$ scaling of the bare spectrum. Instead, they depend on the full nonlinear structure of $\mathcal{F}_I(T,L,\xi)$, which incorporates collective renormalization of the thermodynamic landscape. Once interactions are present, the pressure-like response can be amplified or suppressed not only because the individual levels move with $L$, but also because the collective free-energy functional itself becomes more sharply curved in both temperature and geometry.

\begin{figure}[t]
\centering
\includegraphics[width=0.95\columnwidth]{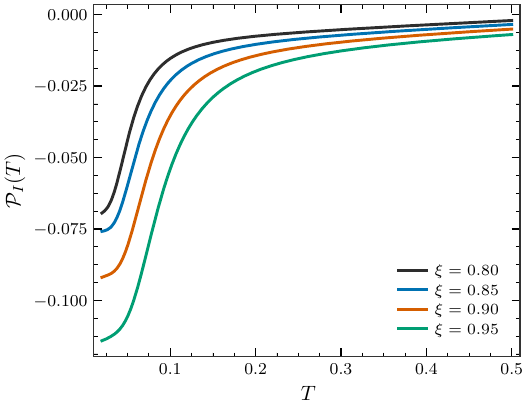}
\caption{Interacting effective pressure $\mathcal{P}_I(T)$ as a function of temperature over the displayed range for $\xi=0.80$, $0.85$, $0.90$, and $0.95$. All curves are negative, indicating that the interacting free energy is lower than in the noninteracting case due to collective exponential resummation. The clear ordering and separation between different couplings demonstrate that the phenomenological interaction preserves and amplifies the $\xi$-dependent spectral structure already present in the bare canonical problem.}
\label{fig:Int-P}
\end{figure}

Figure~\ref{fig:Int-P} presents the interacting effective pressure over the displayed low-to-moderate temperature window. All values of $\mathcal{P}_I$ are negative, which reflects the fact that the exponential resummation in the interacting free energy lowers the total energy relative to the noninteracting case. The smooth ordering of the curves with $\xi$, more negative pressures for larger coupling, confirms that the phenomenological interaction preserves the hierarchy established by the bare spectral deformation while amplifying the magnitude of the response.
\begin{figure}[t]
\centering
\includegraphics[width=0.95\columnwidth]{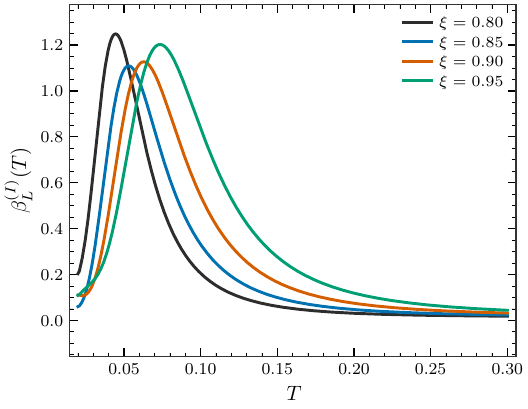}
\caption{Interacting thermal pressure coefficient $\beta_L^{(I)}(T)$ as a function of temperature over the displayed range for $\xi=0.80$, $0.85$, $0.90$, and $0.95$. The sharp, well-separated peaks demonstrate that the phenomenological interaction concentrates the pressure's thermal sensitivity into a narrow, coupling-dependent temperature window. The peak temperature shifts systematically with $\xi$, directly tracking the energy scale at which the corresponding split branches become thermally active in the interacting model.}
\label{fig:Int-beta}
\end{figure}

The interacting thermal pressure coefficient, shown in Fig.~\ref{fig:Int-beta}, provides the most striking illustration of collective amplification in the thermomechanical sector. While the noninteracting canonical $\beta_L$ displays broad, modest peaks, the interacting coefficient develops sharp, well-separated maxima whose positions shift systematically with $\xi$. The peak for $\xi=0.8$ appears at the lowest temperature and has the largest amplitude, reflecting the fact that this coupling places the lowest split levels closest together, so that even a small temperature increase triggers a rapid redistribution of occupation. As $\xi$ increases, the effective gap between the relevant levels grows, and the peak shifts to higher temperature while broadening. This behavior is a direct consequence of the nonlinear resummation encoded in the interacting free energy: the exponential-resummation structure concentrates the thermal response into a narrow window, converting the broad canonical features into sharp, easily detectable signatures.

\begin{figure}[t]
\centering
\includegraphics[width=0.95\columnwidth]{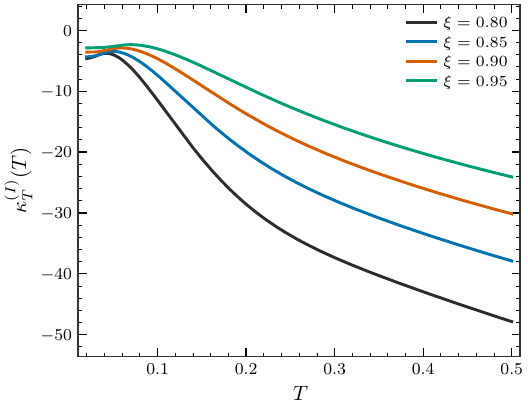}
\caption{Interacting isothermal compressibility $\kappa_T^{(I)}(T)$ as a function of temperature over the displayed range for $\xi=0.80$, $0.85$, $0.90$, and $0.95$. Unlike the grand-canonical case, no divergences appear: the interacting compressibility is everywhere finite and smooth, indicating that the collective exponential resummation regularizes the mechanical response. The ordering of the curves with $\xi$ reflects the coupling-dependent curvature of the interacting free-energy surface with respect to the circumference.}
\label{fig:Int-kappa}
\end{figure}

The interacting isothermal compressibility, shown in Fig.~\ref{fig:Int-kappa}, exhibits qualitatively different behavior from both the canonical and grand-canonical cases. Most notably, no divergences appear: the interacting compressibility is everywhere finite and smooth, indicating that the collective resummation regularizes the mechanical response by preventing the denominator of $\kappa_T^{(I)}$ from passing through zero. The curves are negative throughout the displayed range and are ordered by coupling strength, with stronger $\xi$ yielding smaller absolute values (less compressible systems). This ordering reflects the fact that larger coupling deepens the effective potential well of the interacting free energy, making the system stiffer against geometric deformations. The nonmonotonic temperature dependence visible for $\xi=0.8$ --- where $|\kappa_T^{(I)}|$ first increases before decreasing --- is a fingerprint of the same thermal activation mechanism that produces the sharp $\beta_L^{(I)}$ peak at the same coupling.

\begin{figure}[t]
\centering
\includegraphics[width=0.95\columnwidth]{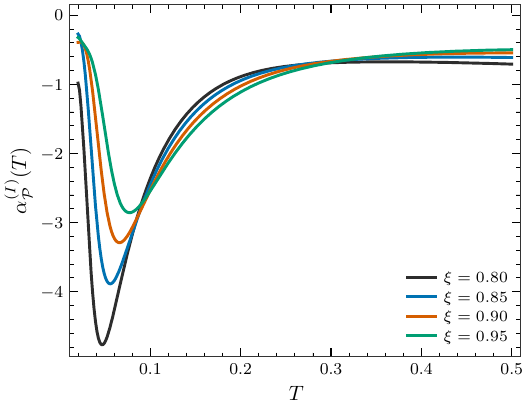}
\caption{Interacting isobaric expansion coefficient $\alpha_{\mathcal{P}}^{(I)}(T)=\kappa_T^{(I)}\beta_L^{(I)}$ as a function of temperature over the displayed range for $\xi=0.80$, $0.85$, $0.90$, and $0.95$. The coefficient is negative and exhibits well-separated minima that shift systematically with the coupling, reflecting the combined action of the sharp thermal-pressure peaks and the smooth compressibility background. The interacting model therefore predicts a regime of anomalous contraction, the system contracts upon heating, that is entirely absent in the noninteracting canonical description, and whose temperature scale is controlled by the effective spin-orbit coupling.}
\label{fig:Int-alpha}
\end{figure}

The interacting expansion coefficient, shown in Fig.~\ref{fig:Int-alpha}, is the most informative synthetic observable of the interacting thermomechanical sector. The coefficient is negative throughout the displayed range, implying that, within this phenomenological model, the ring system contracts upon heating, a form of anomalous thermal expansion driven by the collective reshaping of the free-energy landscape. The well-separated minima shift systematically with $\xi$, tracking the same energy scales identified by the $\beta_L^{(I)}$ peaks. The depth of the minima is largest for $\xi=0.8$, where the coupling places the relevant split levels closest to the thermal activation threshold, and decreases progressively for larger couplings.

This anomalous contraction is physically intuitive within the interacting framework: the exponential resummation makes the interacting free energy more sharply curved than its noninteracting counterpart, and the thermal activation of new levels in such a landscape produces a net inward geometric response rather than the conventional outward expansion. The effect is entirely absent in the noninteracting canonical description and represents a genuine prediction of the interacting model.

Overall, the thermomechanical analysis presented in this section demonstrates that the quantum ring system admits a coherent and physically meaningful set of effective mechanical observables across all three descriptions, canonical, grand canonical, and interacting. The key insight is that geometric confinement is not merely a passive boundary condition: it actively determines the structure of the spectrum and, through it, generates measurable macroscopic response functions. The effective coupling $\xi$ controls the deformation of the angular-momentum branches, and this deformation propagates from the level structure to the pressure-like observable, the compressibility, the thermal pressure coefficient, and the expansion coefficient. The grand-canonical sector amplifies these signatures through Fermi-surface effects, while the interacting extension introduces collective regularization and anomalous contraction. The resulting framework therefore provides a unified description of how geometry, spin--orbit physics, quantum statistics, and collective effects combine to shape the emergent thermomechanical behavior of the quantum ring.

\section{Connection to nonminimal couplings and experimental bounds}
\label{Sec:Bounds}

The dimensionless parameter $\xi = mr_0\mathcal{F}_{12}$ links the macroscopic thermomechanical response developed in this work directly to the underlying nonminimal coupling constants of the relativistic theory. In our companion work~\cite{SilvaReisLisboa2025}, we derived explicit order-of-magnitude bounds on the background tensor field $\mathcal{F}_{12}$ from spectral, transport, and phase-effect observables accessible in semiconductor quantum ring experiments. For representative ring radii $r_0\sim 100\,$nm and effective masses $m\sim 0.067\,m_e$ (typical of InAs rings~\cite{Lorke2000,Fuhrer2001}), those bounds translate into a range of $\xi$ values that encompasses the coupling strengths explored throughout the present paper.

The thermomechanical observables introduced here provide, in principle, additional and potentially sharper probes of $\xi$. In particular, the divergences in the isothermal compressibility $\kappa_T$ and the isobaric expansion coefficient $\alpha_{\mathcal{P}}$ in the grand-canonical ensemble (Sec.~\ref{Sec:ThermomechGrandCanonical}) occur at temperatures that are directly set by the coupling-dependent level-crossing energies. The precise location of these mechanical instabilities as a function of temperature could therefore be used, in conjunction with low-temperature specific-heat or thermal-expansion measurements on ring arrays, to place tight constraints on $\xi$ and, through it, on the nonminimal coupling $\mathcal{F}_{12}$. A quantitative extraction of such bounds from realistic material parameters is left for future work.

\section{Discussion and perspectives}
\label{Sec:Discussion}

The central question addressed by this work is not only whether a Rashba-like Hamiltonian can be engineered from nonminimal couplings, but also what new mesoscopic phenomena become visible once that mechanism is embedded in a ring geometry. The results presented in the preceding sections show that the answer involves a rich interplay among spectral deformation, quantum statistics, and geometric confinement, which manifests in a wide range of thermodynamic and mechanical observables.

A first key observation concerns the role played by the ring geometry in defining thermodynamic and mechanical properties. Unlike conventional thermodynamic systems, where volume is a dynamical variable associated with macroscopic deformations, the radius $r_0$ of the quantum ring appears in the problem as a geometric parameter that defines the underlying Hamiltonian~\cite{Viefers2004,Imry2002}. By adopting a quasi-static interpretation of geometric variations, in which the circumference $L=2\pi r_0$ is promoted to an effective thermodynamic variable, we have shown that variations of $L$ can be understood as infinitesimal, adiabatic deformations that preserve the spectral structure and do not induce transitions between eigenstates. This perspective closely parallels the treatment of lattice parameters in crystalline systems, where thermodynamic response functions are derived from band structures computed at fixed geometry~\cite{PathriaBeal2011,LandauStat}.

Within this framework, the emergence of an effective pressure-like quantity $\mathcal{P}$ should be interpreted as a generalized mechanical response associated with the geometric dependence of the energy spectrum. Rather than describing a physical force in a dynamical sense, $\mathcal{P}$ quantifies how the free energy responds to changes in system size. The associated coefficients, isothermal compressibility, thermal pressure coefficient, and expansion coefficient, encode how the spectral structure reorganizes under infinitesimal geometric deformations.

A central result is that these thermomechanical quantities are directly controlled by the effective coupling~$\xi$, which modifies the spectrum in a nontrivial way by shifting the centers of the angular-momentum branches and altering the ordering of low-energy states~\cite{SilvaReisLisboa2025}. This induces measurable changes in macroscopic observables and establishes the thermomechanical response as an indirect but powerful probe of the underlying spin-orbit structure.

The canonical analysis (Figs.~\ref{fig:FPlot}--\ref{fig:JThermal}) demonstrates that even a single electron per ring is sufficient to produce clear thermodynamic signatures of the spectral reconstruction. The heat capacity, in particular, serves as a spectroscopic tool: the Schottky-like peak splits and shifts to lower temperatures as $\xi$ increases, directly tracking the opening of new low-energy scales due to the Rashba-like interaction. The entropy and internal energy confirm this picture by showing that the thermal activation of excited states occurs at coupling-dependent temperatures.

The grand-canonical analysis (Figs.~\ref{fig:SPlotC2}--\ref{fig:JZgrand}) reveals a qualitatively richer behavior. The Pauli exclusion principle amplifies the effect of spectral shifts near the Fermi edge, producing strong oscillations in magnetization and susceptibility that constitute a mesoscopic analog of the de~Haas--van~Alphen effect~\cite{ByersYang1961,AmbegaokarEckern1990}, but generated here by spin--orbit coupling rather than an external magnetic field. In the classical de~Haas--van~Alphen effect, oscillatory responses occur as a function of $1/B$; here, the coupling parameter $\xi$ plays the analogous role, with its variation driving rearrangements of the near-Fermi-edge occupation in a formally equivalent manner. The sign reversal of the persistent spin current, observed in Fig.~\ref{fig:JZgrand}, is a striking illustration of this mechanism: it arises because thermally activated states above the chemical potential carry spin current opposite to the contribution of the filled states below, an effect that has no analog in the single-particle canonical ensemble.

The thermomechanical analysis developed in Sec.~\ref{Sec:ThermomechanicalResponse} extends this picture to the mechanical domain. In the canonical ensemble, the response is smooth and directly governed by the internal energy density and the heat capacity through the exact relations $\mathcal{P}=2U/L$ and $\beta_L=2C/L$. In the grand-canonical ensemble, Fermi-surface effects generate coupling-dependent mechanical instabilities, divergences in $\kappa_T$ at specific temperatures, that have no counterpart in the canonical case and provide a uniquely sensitive probe of the near-Fermi-edge level structure. The direct comparison in Figs.~\ref{fig:Compare-beta} and~\ref{fig:Compare-alpha} makes this amplification explicit, showing that the grand-canonical response exceeds the canonical one by an order of magnitude and that the expansion coefficient diverges sharply at the instability temperature while its canonical counterpart remains bounded and nearly featureless.

The phenomenological interacting extension (Sec.~\ref{Sec:ThermomechInteracting}) demonstrates that even weak inter-ring couplings can produce dramatic modifications of the thermomechanical response at low temperatures, including anomalous thermal contraction. While a fully microscopic interacting treatment---derived directly from the underlying relativistic theory---remains an important open problem, the present phenomenological approach already illustrates that collective effects in spin--orbit-active ring arrays can qualitatively reshape macroscopic observables.

From a broader perspective, the results presented here demonstrate that antisymmetric tensor fields and their derivatives can induce effective spin--orbit interactions with clear condensed-matter analogues~\cite{SilvaReisLisboa2025,BelichSilva2011,BakkeSilva2012,ReisSilva2024}. This opens several promising directions for future research:

\begin{itemize}
\item[(i)] \emph{Non-equilibrium dynamics.}
  Time-dependent configurations of the background field could generate driven spin precession, geometric-phase pumping, or transitions between spin branches. Studying the corresponding Floquet spectrum may reveal additional topological features.

\item[(ii)] \emph{Disorder and stochastic backgrounds.}
  The dependence of the spin currents on $\varphi$ suggests that noisy or spatially fluctuating backgrounds may produce dephasing and suppression of persistent currents, or alternatively enhance them depending on the disorder correlator.

\item[(iii)] \emph{Extensions to many-body and interacting fermion systems.}
  Arrays of rings or higher-dimensional networks could exhibit collective transport phenomena, spin--orbit--induced phase transitions, or synchronization of spin textures.

\item[(iv)] \emph{Experimental analogues.}
  Although the mechanism is field-theoretic, similar effective Hamiltonians can be engineered in semiconductor quantum rings~\cite{Lorke2000,Fuhrer2001,Nagasawa2013}, cold-atom platforms with synthetic gauge fields~\cite{LinSpielman2011,DalibardRMP2011,GoldmanRPP2014,GalitskiSpielman2013}, or photonic ring resonators where Rashba-like interactions can be tuned dynamically.

\item[(v)] \emph{Coupling to gauge fields or axion-like backgrounds.}
  Extending the model to include dynamical gauge potentials or axion couplings may reveal novel topological responses or spin--orbit--induced anomaly-like features~\cite{CarrollFieldJackiw1990}.
\end{itemize}

Taken together, these perspectives indicate that the Rashba-like interaction induced by antisymmetric-tensor backgrounds is not only theoretically appealing but also opens a pathway to a wide variety of applications in both high-energy and condensed-matter contexts.

\section{Conclusion}
\label{Sec:Conclusion}

In this work, we have developed a comprehensive thermodynamic
and thermomechanical description of a fermionic system confined to a
quantum ring and subject to effective spin--orbit interactions induced
by nonminimal couplings to the antisymmetric tensor
fields~\cite{SilvaReisLisboa2025}. Starting from the exact energy
spectrum, we constructed both canonical and grand-canonical formulations
and extended the analysis to include a phenomenological interacting model
as well as a full thermomechanical framework.

A central result of this study is that the thermodynamic behavior of the system is governed by a nontrivial deformation of the energy spectrum controlled by the effective coupling $\xi$. Unlike simple spin-splitting mechanisms~\cite{BychkovRashba1984,Dresselhaus1955}, this deformation reorganizes the ordering of the energy levels, leading to distinctive thermal signatures that are directly reflected in quantities such as the internal energy, entropy, and heat capacity. The canonical heat capacity, in particular, develops a characteristic multi-peak structure whose evolution with $\xi$ provides a direct
spectroscopic fingerprint of the Rashba-induced spectral reconstruction.

In the grand-canonical ensemble, the Pauli exclusion principle amplifies these signatures considerably. The competition between the coupling-dependent spectrum and Fermi-edge physics produces nonmonotonic magnetization profiles, strongly enhanced susceptibilities, and, most strikingly, sign reversals in the persistent spin current driven by the thermal redistribution of occupied branches.
These features constitute a mesoscopic analog of the de~Haas--van~Alphen
effect~\cite{ByersYang1961}, generated entirely by spin-orbit coupling
rather than an external magnetic field: the coupling $\xi$ plays the role
of the inverse applied field, and the near-Fermi-edge occupation
oscillates accordingly as $\xi$ is varied.

Building on this spectral structure, we introduced a consistent
thermomechanical framework by promoting the ring circumference
$L=2\pi r_0$ to an effective thermodynamic variable.
Within a quasi-static interpretation, geometric variations are treated
as parametric deformations of the spectrum, allowing mechanical response
functions to be defined without altering the underlying eigenstates.
This approach led to the identification of an effective pressure-like
quantity and the corresponding response coefficients, including the
isothermal compressibility~$\kappa_T$, the thermal pressure
coefficient~$\beta_L$, and the isobaric expansion
coefficient~$\alpha_{\mathcal{P}}$.

An important conceptual outcome is that these thermomechanical
quantities are not introduced phenomenologically, but instead emerge
directly from the quantum structure of the system.
In particular, we have shown that the effective pressure satisfies
$\mathcal{P}=2U/L$, while the thermal pressure coefficient obeys
$\beta_L=2C/L$. These exact relations provide a transparent
connection between microscopic spectral properties and macroscopic
response functions, establishing the geometric confinement not merely
as a constraint, but as a fundamental ingredient that enables the
emergence of effective thermomechanical behavior.

In the grand-canonical sector, Fermi-surface effects amplify the
thermomechanical response dramatically, producing coupling-dependent
divergences in the compressibility and expansion coefficient that signal
mechanical instabilities of the confined quantum system.
These features are entirely absent in the canonical ensemble, as
demonstrated explicitly by the direct comparison in
Figs.~\ref{fig:Compare-beta} and~\ref{fig:Compare-alpha}, which show that the grand-canonical response exceeds its canonical counterpart by an order of magnitude and develops sharp divergent structures that the canonical ensemble completely lacks.

The extension to the phenomenological interacting model demonstrates
that even weak collective effects can amplify the thermodynamic response
by more than an order of magnitude, concentrating the sensitivity to
the spin--orbit parameter into a narrow low-temperature window.
Remarkably, the interacting expansion coefficient is negative
throughout the low-temperature regime, predicting anomalous thermal
contraction driven by collective reshaping of the free-energy landscape.
This suggests that weakly coupled ring arrays could serve as sensitive
probes of spin-orbit interactions and points toward the possibility of
engineering systems with tailored thermomechanical properties through
controlled inter-ring coupling.

Overall, the framework developed in this work provides a unified
description of spectral, thermodynamic, and mechanical properties in a
geometrically constrained quantum system.
It opens the possibility of exploring similar mechanisms in a broad
class of mesoscopic and engineered systems, including topological
rings~\cite{HasanKane2010,QiZhang2011}, cold-atom platforms with
synthetic spin--orbit
coupling~\cite{LinSpielman2011,GalitskiSpielman2013}, and photonic ring
resonators~\cite{GoldmanRPP2014}, where geometry and effective gauge
structures play a central role in shaping macroscopic observables.

\section*{Acknowledgments}
This work was partially supported by the Brazilian agencies CAPES, CNPq, FAPEMA and FAPESB. EOS acknowledges the support from grants CNPq/306308/2022-3,
FAPEMA/UNIVERSAL-06395/22, and CAPES/Code 001. J.A.A.S.R acknowledges partial financial support from UESB through Grant AuxPPI (Edital No. 267/2024), as well as from FAPESB--CNPq/Produtividade under Grant No. 12243/2025 (TOB-BOL2798/2025).

\end{document}